\documentclass[letterpaper,11pt]{article}
\pdfoutput=1

\usepackage{jheppub}
\usepackage{multirow}
\usepackage{subfig}
\usepackage{xspace}
\usepackage{xcolor}
\usepackage[countmax]{subfloat}
\usepackage{amsmath}
\usepackage{mathtools}
\usepackage{graphicx}
\usepackage{slashed}
\usepackage[normalem]{ulem}
\usepackage{tikz-feynman,contour}
\usepackage{tikz,pgf}
\usepackage{braket}
\usepackage{amsfonts}
\usepackage{amsthm,scrextend,bbold}
\usepackage{comment}
\usepackage{appendix}
\usepackage{subfiles}

\newcommand{\as}{\alpha_\text{s}}
\newcommand{\cf}{C_{\text{F}}}
\newcommand{\ca}{C_{\text{A}}}
\newcommand{\tr}{T_{\text{R}}}

\newcommand{\order}[1]{{\cal O}\left(#1\right)}

\DeclareMathOperator{\De}{d}
\newcommand{\de}{\De\!}

\newcommand{\kt}{k_{\text{T}}}
\newcommand{\mt}{m_{\text{T}}}

\newcommand{\gs}{\gamma_\text{soft}}
\newcommand{\gszero}{\gamma_\text{soft}^{(0)}}

\newcommand{\gsone}{\gamma_\text{soft}^{(1)}}

\newcommand{\ord}[1]{{\cal O}\left(#1\right)}
\usepackage{color}
\definecolor{darkblue}{rgb}{0,0,0.5}
\definecolor{darkgreen}{rgb}{0,0.5,0}
\definecolor{darkorange}{rgb}{0.8,0.3,0}

\DeclareRobustCommand{\Sec}[1]{Sect.~\ref{#1}}

\newcommand{\pythia}[1]{\textsc{Pythia\xspace #1}}

\newcommand{\herwig}[1]{\textsc{Herwig\xspace #1}}

\newcommand{\vincia}[1]{\textsc{Vincia\xspace #1}}

\newcommand{\mathematica}{\texttt{Mathematica }}

\preprint{  
\parbox[t]{40 mm}{CERN-TH-2026-273\\ TUM-HEP-1619/26 }}
\title{
Soft-gluon effective coupling for massive quarks}
\author[1]{Luca Bason,}
\author[2]{Simone Devoto,}
\author[3]{Andrea Ghira,}
\author[1]{Simone Marzani,}
\author[4]{Pier Francesco Monni,}
\author[1]{and Giovanni Ridolfi}

\affiliation[1]{Dipartimento di Fisica, Universit\`a di Genova and INFN, Sezione di Genova, Via Dodecaneso 33, 16146, Italy}
\affiliation[2]{Vakgroep Fysica en Sterrenkunde, Universiteit Gent, 9000 Gent, Belgium}
\affiliation[3]{Technical University of Munich, TUM School of Natural Sciences, Physics Department, James-Franck-Straße 1, 85748 Garching, Germany}
\affiliation[4]{CERN, Theoretical Physics Department, CH-1211 Geneva 23, Switzerland}

\emailAdd{luca.bason@gmail.com}
\emailAdd{simone.devoto@UGent.be}
\emailAdd{andrea.ghira@tum.de}
\emailAdd{simone.marzani@ge.infn.it}
\emailAdd{pier.monni@cern.ch}
\emailAdd{giovanni.ridolfi@ge.infn.it}

\abstract{
We present a generalisation of the soft-gluon physical coupling, defined in terms of the strength of soft radiation from a hard dipole, to massive hard emitters. We derive the relation between this coupling and the $\overline{\rm MS}$ coupling through two loops, encompassing all configurations in which both, one, or neither of the hard emitters are massive. We show that the presence of massive emitters introduces additional freedom in the definition of the physical coupling scheme, which we characterise at two-loop order. Our results provide new ingredients for precision QCD resummation in processes involving heavy quarks and for the development of higher-order parton-shower algorithms.
}

\begin{document}
\maketitle

\newpage

\section{Introduction}\label{sec:intro}
Perturbative predictions for hard-scattering observables develop large
logarithms when the measured final state approaches an exclusive boundary of
phase space.  In this region, real radiation is forced to be soft or collinear,
and its incomplete cancellation against virtual corrections generates Sudakov
logarithms.  For a broad class of observables these logarithms exponentiate and
can be organised in a hierarchy of logarithmic accuracies, see for instance~\cite{Luisoni:2015xha} and references therein.
%
The same soft and collinear limits are the basic ingredients of parton-shower algorithms.  Controlling the
inclusive rate and the kinematics of unresolved radiation is therefore central
both to analytic resummation and to the construction of showers with stated
logarithmic accuracy~\cite{Dasgupta:2018nvj,Dasgupta:2020fwr,vanBeekveld:2022ukn,Forshaw:2020wrq,Nagy:2020dvz,Nagy:2020rmk,Herren:2022jej,Assi:2023rbu,Preuss:2024vyu,Hoche:2024dee,vanBeekveld:2024wws,vanBeekveld:2025lpz,vanBeekveld:2026wnj}.

For radiation from massless hard partons, the inclusive effect of a soft
gluon splitting can be absorbed, through next-to-leading logarithmic (NLL)
accuracy, into the Catani--Marchesini--Webber (CMW) coupling
\cite{Catani:1990rr}.  The CMW prescription replaces the
$\overline{\mathrm{MS}}$ coupling in the one-gluon emission probability by a
physical coupling that includes the correlated decay of the emitted gluon.
This construction is universal at this order: it encapsulates the only genuine two-loop 
ingredient for global resummations (for instance threshold, transverse-momentum and event-shape resummation), 
and it describes the effective next-to-leading order (NLO) single-emission probability entering NLL parton showers at leading-colour.

Extending this idea beyond NLL accuracy requires a more precise definition of
what is meant by an inclusive soft emission.  A useful formulation starts from
the infrared- and collinear-safe probability density for correlated soft
radiation with total momentum $k$.  Integrating this density while holding a
chosen function of the transverse momentum and transverse mass fixed defines a
family of soft-gluon effective couplings
\cite{Banfi:2018mcq,Catani:2019rvy,Catani:2023kpw}.  
All members of the family
reduce to the CMW coupling through $\mathcal{O}(\as^2)$ in four dimensions, but
they generally differ at higher orders.  Their equality at the conformal Wilson-Fisher fixed point
and their relation to the light-like cusp anomalous dimension provide an
important structural constraint on these definitions~\cite{Catani:2019rvy,Catani:2023kpw}.
The generalization of the physical coupling scheme beyond the soft limit 
has also been investigated in Refs.~\cite{Dasgupta:2021hbh,vanBeekveld:2023lsa,vanBeekveld:2024qxs}.

Over the past few years, the number of investigations devoted to the jet substructure of heavy-flavour jets has increased substantially~\cite{Maltoni:2016ays, Lee:2019lge,Llorente:2014bha,Li:2017wwc,Li:2021gjw, Craft:2022kdo,Cunqueiro:2022svx,Fedkevych:2022mid,Caletti:2023spr,Blok:2023ugf,Zhang:2023jpe,Dhani:2024gtx,Dhani:2025fbk,Ghira:2026vzw,Ghira:2025nym}.
Among the most prominent QCD effects influencing heavy-flavour jet substructure is the so-called dead-cone effect~\cite{Dokshitzer:1991fd,Dokshitzer:1995ev}, namely the suppression of collinear radiation around massive quarks.
In parallel, parton-shower treatments of massive emitters have recently undergone substantial development.  
The PanScales collaboration has formulated final-state showers
with massive quarks and demonstrated their NLL accuracy through fixed-order and
all-order tests~\cite{vanBeekveld:2026wnj}.  In the ALARIC framework, massive
final-state evolution is based on massive kinematics and azimuthally dependent
splitting kernels that reproduce the soft-radiation pattern and admit fully
differential NLO matching~\cite{Assi:2023rbu}.  General-purpose generators provide
complementary implementations: \pythia{8.3}, and in particular its \vincia{}
antenna shower, systematically includes final-state mass effects and the relevant
quasi-collinear limits~\cite{Bierlich:2022pfr}, while \herwig{7} incorporates the
heavy-quark dead cone in its angular-ordered shower and massive-emitter kinematics
in its dipole shower~\cite{Cormier:2018tog,Bellm:2025pcw}.

Although it has been observed~\cite{Ghira:2023bxr} that the usual CMW coupling suffices to achieve NLL accuracy for the resummation of standard observables even in the presence of massive quarks, a general derivation of this result is still lacking. A systematic analysis of the soft-gluon coupling for massive emitters is therefore needed, both to establish the scope of this observation and to determine how the coupling should be generalised for NNLL resummation. 
In this context, the main complication one has to face stems from the fact that the massive soft (eikonal) current depends on additional dimensionless ratios built
from the masses, the hard momenta and the total soft momentum. 
Consequently,
an inclusive integration over the invariant mass of a correlated soft system
must specify which longitudinal variable is kept fixed.  Choices based on a
rapidity-like variable and choices based on a momentum fraction agree in the
massless limit, but need not define the same effective coupling at finite mass.
This distinction is relevant whenever one seeks a common description of soft
radiation in analytic resummation and in heavy-quark parton showers.

In this work, we construct the soft correlated-emission density for a dipole of
massive quarks through $\mathcal{O}(\as^2)$.  We combine the one-loop massive
soft current with the tree-level emission of two gluons or a light
quark--antiquark pair, using the massive soft currents collected in
Refs.~\cite{Czakon:2014oma,Catani:2023tby}.  We then integrate over the invariant
mass of the unresolved system and study how the result depends on the
kinematic variable held fixed. This exposes which part of the CMW correction
remains universal for massive emitters and which part belongs to the choice of
soft-coupling scheme.

The paper is organised as follows.  In \Sec{sec:cmw} we review the definition
of the correlated-emission density and the massless soft-gluon coupling.  In
\Sec{sec:massive_emitters} we formulate and evaluate the corresponding density
for massive emitters.  We discuss the resulting massive coupling schemes and
their applications in Sect.~\ref{sec:pheno}, before drawing our conclusions in Sec.~\ref{sec:conclusions}. Technical details of the real-radiation
integrals are collected in the appendices.

\section{Soft-gluon coupling}\label{sec:cmw}
In this section we review the construction of the soft-gluon effective
coupling for two massless emitters.  
This fixes our notation and conventions and sets the stage
for the non-trivial generalisation to massive quarks.

\subsection{Recap of the massless NLO calculation}
\label{sec:massless_nlo_sketch}

Let us consider a generic process containing two massless hard partons with momenta $p_1$ and $p_2$, possibly together with other colour-singlet particles and $N$ additional soft partons.
When their momenta $k_1,\ldots,k_N$ become uniformly
soft, the squared matrix element factorises to all orders in the loop expansion~\cite{Catani:1999ss,Catani:2000pi} as
\begin{equation}
 \left|\mathcal{M}_N(p_1,p_2,k_1,\ldots,k_N)\right|^2
 \simeq
 \left|\mathcal{M}_0(p_1,p_2)\right|^2
\left|J_i(k_1,\ldots,k_N)\right|^2 .
\label{eq:soft_factorisation_massless}
\end{equation}
The soft current $J_i$ ($i=q,g$) contains the complete leading-power dependence on the
soft momenta.  
Let us introduce the total soft momentum:
\begin{equation}
 k=\sum_{\ell=1}^{N}k_\ell,
 \label{eq:total_soft_momentum}
\end{equation}
and the following kinematic invariants
\begin{equation}
 \mt^2=\frac{2(p_1\!\cdot k)(p_2\!\cdot k)}{p_1\!\cdot p_2},
 \qquad
 \kt^2=\mt^2-k^2,
 \qquad
 \hat{y}=\frac{1}{2}\ln\frac{p_2\!\cdot k}{p_1\!\cdot k}.
 \label{eq:massless_soft_variables}
\end{equation}
In these variables the invariant phase-space integrated over the $d-3$ transverse angles reads
\begin{equation}
 \de^d k=\frac{1}{2}\,\de \hat{y}\,\de \mt^2\,\de \kt^2\,
 \mathcal{N}_\epsilon(\kt),
 \qquad
 \mathcal{N}_\epsilon(\kt)=
 \frac{\pi^{1-\epsilon}(\kt^2)^{-\epsilon}}{\Gamma(1-\epsilon)},
 \label{eq:massless_phase_space_jacobian}
\end{equation}
where we have set $d=4-2 \epsilon$.

In order to define the soft-gluon effective coupling~\cite{Banfi:2018mcq,Catani:2019rvy}, we consider the probability of correlated emission (including the corresponding virtual corrections) of an arbitrary number of soft partons with total momentum $k$
\begin{equation} \label{eq:prob-density}
    \de \mathcal{S}_i= \frac{1}{2}\,\de \hat{y}\,\de \mt^2\,\de \kt^2\ w_i(k;\epsilon),
\end{equation}
which is expressed in terms of the web function~\cite{Gatheral:1983cz,Frenkel:1984pz}:
\begin{align}
w_i(k;\epsilon)=\mathcal{N}_\epsilon(\kt)\Bigg\{&
 \frac{\delta_+(k^2)}{(2\pi)^{d-1}}\left|J_i(k)\right|^2
 \nonumber\\
 &+\sum_{N=2}^{\infty}\ \sum_{\{a_1,\ldots,a_N\}}
 \sigma_{a_1\ldots a_N}
 \int\!\left[\prod_{\ell=1}^{N}[\de k_\ell]\right]
 \nonumber\\[-0.3ex]
 &\hspace{2.1cm}\times
 W_i^{a_1\ldots a_N}(k_1,\ldots,k_N)
 \delta^{(d)}\!\left(k-\sum_{\ell=1}^{N}k_\ell\right)
 \Bigg\},
 \label{eq:web_all_order}
\end{align}
where
\begin{equation}
 [\de k_\ell]=\frac{\de^d k_\ell}{(2\pi)^{d-1}}
 \delta_+(k_\ell^2).
 \label{eq:soft_phase_space_measure}
\end{equation}
In Eq.~\eqref{eq:web_all_order}, the first term is the soft-gluon current~\cite{Catani:2000pi}, while $a_\ell$ labels the flavour of each soft parton,
$\sigma_{a_1\ldots a_N}$ is the identical-particle symmetry factor, and
$W_i^{a_1\ldots a_N}$ is the irreducible part of the squared soft current:
all products of lower-multiplicity correlations are subtracted.  Both the
single-emission current and the irreducible currents in
Eq.~\eqref{eq:web_all_order} include their virtual corrections.  
Real and virtual singularities
cancel in $w_i$ order by order, so it has a finite four-dimensional limit~\cite{Banfi:2018mcq,Catani:2023kpw}.
Note that the invariance of the massless eikonal current under independent rescalings of
$p_1$ and $p_2$ implies that the correlated-emission density does not depend on rapidity.

Let us now review the calculation of the web function in the case of massless emitters to NLO accuracy:
\begin{equation}
w_i(k;\epsilon)=w_{i,\mathrm{LO}}(k;\epsilon)+w_{i,\mathrm{NLO}}(k;\epsilon)+ \ord{\as^3}.
\end{equation}
At LO, the soft system consists of one on-shell gluon ($k^2=0$), so the squared eikonal current is
\begin{equation}
 \left|J_i(k)\right|^2_{\mathrm{LO}}
 =8\pi\,C_i\,\as(\mu^2)(\mu^2)^\epsilon S_\epsilon^{-1}
   \frac{p_1\!\cdot p_2}{(p_1\!\cdot k)(p_2\!\cdot k)},
 \label{eq:massless_lo_current}
\end{equation}
where $ S_\epsilon=(4\pi)^\epsilon e^{-\epsilon\gamma_E}$ and $\mu$ is the renormalisation scale.
At this order only the first term in Eq.~\eqref{eq:web_all_order} contributes and we have
\begin{equation}
 w_{i,\mathrm{LO}}(k;\epsilon)
 =\mathcal{N}_\epsilon(\kt)
 \frac{\delta_+(k^2)}{(2\pi)^{d-1}}
 \left|J_i(k)\right|^2_{\mathrm{LO}}=
 \frac{2}{\mt^2}\,\,C_i
   \frac{\as(\mu^2)}{\pi}
   \left(\frac{\mu^2}{\kt^2}\right)^\epsilon c(\epsilon) \delta(\mt^2-\kt^2),
 \label{eq:lo_web_from_current}
\end{equation}
with 
\begin{equation}
 c(\epsilon)=\frac{e^{\epsilon\gamma_E}}{\Gamma(1-\epsilon)}.
 \label{eq:dimensional_factors}
\end{equation}

At NLO, i.e.\
$\mathcal{O}(\as^2)$, both virtual and real-emission corrections to soft-gluon emission must be included.
The former is given by the one-loop correction to the soft current~\cite{Catani:2000pi}:
\begin{align}
 \left|J_i(k)\right|^2_{\mathrm{NLO}}
 =-\as(\mu^2)\Bigg[&
 \frac{c(2\epsilon)}{2c(\epsilon)}\ca
 \frac{\pi\cos(\pi\epsilon)}{\sin^2(\pi\epsilon)}
 \left(
 \frac{\mu^2(p_1\!\cdot p_2)}
 {2(p_1\!\cdot k)(p_2\!\cdot k)}
 \right)^\epsilon
+\frac{\beta_0}{\epsilon}\Bigg]
 \left|J_i(k)\right|^2_{\mathrm{LO}},
 \label{eq:massless_one_loop_current}
\end{align}
with $\beta_0=\frac{11\ca-4\tr n_f}{12 \pi}$.
The real-emission contribution can be decomposed as follows:
\begin{subequations}\label{eq:double_soft_correlations}
    \begin{align}
 \left|J_i^{gg}(k_1,k_2)\right|^2_{\mathrm{LO}}
 &=\left|J_i(k_1)\right|^2_{\mathrm{LO}}
   \left|J_i(k_2)\right|^2_{\mathrm{LO}}
   +W_i^{gg}(k_1,k_2), \label{eq:double_soft_correlations_gg}
 \\
 \left|J_i^{q\bar q}(k_1,k_2)\right|^2_{\mathrm{LO}}
 &=W_i^{q\bar q}(k_1,k_2),\label{eq:double_soft_correlations_qq}
\end{align}
\end{subequations}
where the two contributions account for the emission of two soft gluons or a soft quark-antiquark pair, respectively. 
Note that the product in Eq.~\eqref{eq:double_soft_correlations_gg} describes independent emissions, and therefore does not contribute to the connected web.  Thus the $\ord{\as^2}$ contribution to Eq.~\eqref{eq:web_all_order} is
\begin{align}
 w_{i,\mathrm{NLO}}(k;\epsilon)
 =\mathcal{N}_\epsilon(\kt)\Bigg\{&
 \frac{\delta_+(k^2)}{(2\pi)^{d-1}}
 \left|J_i(k)\right|^2_{\mathrm{NLO}}
 \nonumber\\
 &+\int[\de k_1][\de k_2]\,
 \delta^{(d)}(k-k_1-k_2)
 \left[\frac{1}{2}W_i^{gg}(k_1,k_2)
       +n_f W_i^{q\bar q}(k_1,k_2)\right]\Bigg\},
 \label{eq:massless_nlo_web}
\end{align}
where the factor $1/2$ is the Bose symmetry factor for the gluons.  Equation
\eqref{eq:massless_nlo_web} makes the inclusive content of the effective
coupling explicit: a one-loop unresolved gluon is combined with all of its
two-parton real decay channels at fixed total momentum.

We briefly describe the calculation of the real-emission contributions to Eq.~\eqref{eq:massless_nlo_web} to set the stage for the case with massive emitters of the next section.
We start from the quark-antiquark contribution. In this case, it is useful to write it as
\begin{align}
 W_i^{q\bar q}(k_1,k_2)
&=\left(4\pi\as(\mu^2)\mu^{2\epsilon}S_\epsilon^{-1}\right)^2
 C_i\,\widehat J_\mu(k_1+k_2)\,
 \Pi_{q\bar q}^{\mu\nu}(k_1,k_2)\,\widehat J_\nu(k_1+k_2),
 \label{eq:massless_qq_current}
\end{align}
where
\begin{equation}
     \widehat J^\mu(k)
 =\frac{p_1^\mu}{p_1\!\cdot k}-\frac{p_2^\mu}{p_2\!\cdot k},
\end{equation}
and
\begin{equation}
 \Pi_{q\bar q}^{\mu\nu}(k_1,k_2)
 =\frac{\tr}{(k_1\!\cdot k_2)^2}
 \left[-g^{\mu\nu}(k_1\!\cdot k_2)
 +k_1^\mu k_2^\nu+k_1^\nu k_2^\mu\right].
 \label{eq:massless_qq_splitting_tensor}
\end{equation}
Current conservation then implies that the integrated splitting tensor has
the transverse form
\begin{equation}
 \int[\de k_1][\de k_2]\,\delta^{(d)}(k-k_1-k_2)\,
 \Pi_{q\bar q}^{\mu\nu}(k_1,k_2)
 =A_{q\bar q}(k^2;\epsilon)
 \left(-g^{\mu\nu}+\frac{k^\mu k^\nu}{k^2}\right),
 \label{eq:massless_qq_tensor_reduction}
\end{equation}
Taking the trace gives the explicit coefficient
\begin{equation}
 A_{q\bar q}(k^2;\epsilon)=
 \frac{\tr(1-\epsilon)(k^2)^{-1-\epsilon}}
 {(2\pi)^{d-1}16^{1-\epsilon}\pi^{3/2-\epsilon}
  \Gamma(5/2-\epsilon)}.
 \label{eq:massless_qq_tensor_coefficient}
\end{equation}

We now turn to the double-gluon channel.  Starting from the correlated
two-gluon squared current and exploiting colour conservation, we write
\begin{align}
 W_i^{gg}(k_1,k_2)
 &=-\left(4\pi\as(\mu^2)\mu^{2\epsilon}
 S_\epsilon^{-1}\right)^2 \ca
\sum_{r,s=1}^{2}\mathbf{T}_r\!\cdot\mathbf{T}_s\,
 S_{rs}(k_1,k_2)
 \nonumber\\
 &=\left(4\pi\as(\mu^2)\mu^{2\epsilon}
 S_\epsilon^{-1}\right)^2 \ca C_i
 \left[\widehat S(k_1,k_2)+2\widetilde S(k_1,k_2)\right],
 \label{eq:massless_gg_Sij}
\end{align}
where following~\cite{Catani:2023tby,Catani:2023kpw}, we have separated the soft function into a
contribution that has a structure analogous to the $q\bar q$ case,
\begin{equation}
     \widehat S(k_1,k_2)
 =-\widehat J_\mu(k_1+k_2)\widetilde\Pi^{\mu\nu}(k_1,k_2)
   \widehat J_\nu(k_1+k_2),
\end{equation}
where the tensor structure is given by
\begin{equation}
 \widetilde\Pi^{\mu\nu}(k_1,k_2)=
 -\frac{-4g^{\mu\nu}(k_1\!\cdot k_2)
 +(1-\epsilon)(k_1^\mu k_2^\nu+k_1^\nu k_2^\mu)}
 {(k_1\!\cdot k_2)^2},
 \label{eq:massless_gg_polarisation_tensor}
\end{equation}
and a remainder, $\widetilde S(k_1,k_2)$. The explicit expression of the latter is the same as that of the contribution $\widetilde S^{m=0}_{rs}$ given in Appendix~\ref{app:masters}.

The integrated $\widehat{S}$-contribution is obtained directly from
Eq.~\eqref{eq:massless_qq_tensor_coefficient} by the replacement
\begin{equation}
 n_f \tr\longrightarrow
 -\ca\frac{11-7\epsilon}{4(1-\epsilon)},
 \label{eq:massless_gg_replacement}
\end{equation}
including the factor $1/2$ for identical gluons \cite{Catani:2023tby}. 

The calculation of the remaining terms is most-easily performed in the rest frame of the total soft momentum $k=k_1+k_2$,
so that the only non-trivial geometry is parametrised by the angle $\chi$
between the two hard directions after boosting them into this frame:  \begin{equation}
 \cos\chi
 =1-\frac{k^2(p_1\!\cdot p_2)}{(p_1\!\cdot k)(p_2\!\cdot k)}
 =2t-1,
 \label{eq:massless_opening_angle}
\end{equation}
where we have introduced
\begin{equation}
 t=\frac{\kt^2}{\mt^2}=\frac{\kt^2}{k^2+\kt^2},\qquad 0\leq t\leq1.
 \label{eq:t_def}
\end{equation}
The integrals over the soft-gluon momenta can be expressed in terms of three master integrals 
\begin{align} \label{eq:massless_gg_remainder_integrated}
 \int[\de k_1][\de k_2]\,\delta^{(d)}(k-k_1-k_2)\,
\widetilde{S}(k_1,k_2)
 &=
 \frac{\pi^{-11/2+3\epsilon}}
 {64^{1-\epsilon}\Gamma(1/2-\epsilon)}
 \frac{(k^2)^{-1-\epsilon}(p_1\!\cdot p_2)}
 {(p_1\!\cdot k)(p_2\!\cdot k)}
 \\
 &\times\left[
 2t\,I_{11}^{+}(\chi;\epsilon)
 -4(1-t)I_{11}^{-}(\chi;\epsilon)
 +2I_{10}(\chi;\epsilon)\right],\nonumber
\end{align}
where
\begin{equation}
 I_{ab}^{\pm}(\chi;\epsilon)=
\int_0^\pi\!\de\theta\int_0^\pi\!\de\phi\,
 \frac{(\sin\theta)^{1-2\epsilon}(\sin\phi)^{-2\epsilon}}
 {(1-\cos\theta)^a
  (1\pm\cos\chi\cos\theta
    \pm\sin\chi\sin\theta\cos\phi)^b}.
 \label{eq:massless_angular_masters}
\end{equation}
For $b=0$ the sign is immaterial, and we denote either choice by $I_{a0}$.
The master integrals evaluate to Euler gamma functions and hypergeometric functions of $t$~\cite{Catani:2023tby,Catani:2023kpw}.
Combining these real contributions with the one-loop current gives the $\ord{\as^2}$
$d$-dimensional result as a function of the dimensionless ratio $t$~\cite{Catani:2023kpw}:
\begin{align}
 w_{i,\mathrm{NLO}}(k;\epsilon)
 &=C_i\left(\frac{\as(\mu^2)}{\pi}\right)^2
 \left(\frac{\mu^2}{\kt^2}\right)^{2\epsilon}
 \frac{c(2\epsilon)}{\kt^4}
 \Bigg\{
 \nonumber\\
 &\quad
 \left[-\left(\frac{\mu^2}{\kt^2}\right)^{-\epsilon}
 \frac{c(\epsilon)}{c(2\epsilon)}
 \frac{11\ca-4\tr n_f}{6\epsilon}
 -\ca\frac{\pi^2\cos(\pi\epsilon)}{\sin^2(\pi\epsilon)}\right]
 \delta(1-t)
 \nonumber\\
 &-\big[(12-9\epsilon)\ca-4\tr n_f\epsilon\big]
 \frac{(1-\epsilon)^2\Gamma(1-2\epsilon)}
 {\epsilon\Gamma(4-2\epsilon)}
 (1-t)^{-1-\epsilon}t^{2+\epsilon}
 \nonumber\\
 &
+\ca(1-t)^{-\epsilon}t^{2+\epsilon}
 \Bigg[
 \frac{2}{1+\epsilon}{}_2F_1(1,1;2+\epsilon;1-t)
 -\frac{1}{1-\epsilon}{}_2F_1(1,1;2-\epsilon;1-t)
 \Bigg]\nonumber\\ &+\ca\frac{2\pi}{\sin(\pi\epsilon)}
 (1-t)^{-1-2\epsilon}t^{2+2\epsilon}
 \Bigg\}.
\end{align}
 The expansion of the above expression in plus distributions makes the cancellation of the real endpoint singularities
against the virtual poles explicit, and the result has a finite
$\epsilon\to0$ limit: 
\begin{align}
w_{i,\mathrm{NLO}}(k;\epsilon)
=
\frac{C_i}{\kt^4}
\left(\frac{\as(\mu^2)}{\pi}\right)^2 t^2
\Bigg\{
&\left[ - 2\pi\beta_0\ln\frac{\kt^2}{\mu^2}+
 \ca \left(\frac{67}{18}-\zeta_2\right)
 -\frac{10}{9}\tr n_f
\right]\delta(1-t)
\nonumber\\
&
+\left[
\frac{\ca\ln t-2\ca\ln(1-t)-2\pi\beta_0}{1-t}
\right]_+
\Bigg\}
+\ord{\epsilon}.
\end{align}

\subsection{The CMW coupling and its extensions}\label{sec:cmw_extensions}

The definition of the soft physical
coupling involves a prescription for mapping the total (off-shell) momentum $k$ of the soft radiation
onto a massless momentum $p_k$, as discussed, for instance, in Refs.~\cite{Banfi:2004yd,Banfi:2021owj}. 
Different projections distribute the soft-radiation density differently in the Born phase space, and their 
interpretation as coupling schemes requires that their angular distribution in the dipole is identical to that of the leading-order density.
For a massless dipole, the leading-order soft-radiation density is
invariant under longitudinal boosts of the radiation momentum relative to the
fixed dipole ends. 
Consequently, the massless web depends only on the two scales $\kt$ and $\mt$ and not on the rapidity in the dipole frame.
Thus, preserving $\kt$ (or $\mt$) and $\hat{y}$ respects the desired symmetry and yields an angle-independent
rescaling of the leading-order density.~\footnote{Prescriptions conserving quantities that differ from $\kt$ and $y$ by
a $t$-dependent~\eqref{eq:t_def} amount will have the same property, while other projections (for instance preserving pseudo-rapidity)
introduce additional angular dependence. Such prescriptions are therefore not suitable for a definition of the coupling scheme.}
The ($d$-dimensional) effective coupling
is then defined after specifying the transverse scale that is held fixed in the
inclusive integration:
\begin{align} \label{eq:general_soft_coupling}
 \widetilde{\mathcal{A}}_{\mathcal{F},i}
 \bigl(\as;\epsilon\bigr)
&= \mu^2\int_0^\infty\!\de \mt^2\,\de \kt^2\,
 \delta\!\left(\mu^2-\frac{\kt^2}{\mathcal{F}\left(\frac{\kt^2}{\mt^2}\right)}\right) \frac{\de^3 \mathcal{S}_i}{\de \kt^2 \de \mt^2 \de \hat{y}}\nonumber\\
 &=\frac{1}{2}\mu^2\int_0^\infty\!\de \mt^2\,\de \kt^2\,
 \delta\!\left(\mu^2-\frac{\kt^2}{\mathcal{F}\left(\frac{\kt^2}{\mt^2}\right)}\right)
 w_i(k;\epsilon),
\end{align}
where $\mathcal{F}(t)$ is smooth and is normalised by
$\mathcal{F}(1)=1$~\cite{Catani:2023kpw}.  Two useful choices are
\begin{equation}
 \mathcal{F}\left(t\right)=1;
 \qquad
 \mathcal{F}\left(t\right)=t,
 \label{eq:AT_A0_choices}
\end{equation}
which define the transverse-momentum coupling
$\widetilde{\mathcal{A}}_{\text{T},i}$ and the transverse-mass coupling
$\widetilde{\mathcal{A}}_{0,i}$, respectively.  The physical
four-dimensional coupling is
\begin{equation}
 \mathcal{A}_{\mathcal{F},i}(\as)
 =\widetilde{\mathcal{A}}_{\mathcal{F},i}(\as;0)
 =\sum_{n=1}^{\infty}\left(\frac{\as}{\pi}\right)^n
  \mathcal{A}^{(n)}_{\mathcal{F},i}.
 \label{eq:soft_coupling_expansion}
\end{equation}

At LO, we have
$\mathcal{A}^{(1)}_{\mathcal{F},i}=C_i$ for every admissible
$\mathcal{F}$.  More remarkably, the independence of $\mathcal{F}$ persists in
four dimensions at the next order.  All the couplings in
Eq.~\eqref{eq:general_soft_coupling} reduce through
$\mathcal{O}(\as^2)$ to the CMW radiation intensity,
\begin{subequations}\label{eq:cmw-definition}
\begin{align}
 \mathcal{A}^{\mathrm{CMW}}_i(\as)
 &=C_i\frac{\as}{\pi}
 \left[1+\frac{\as}{2\pi}K\right]
 +\mathcal{O}(\as^3),
 \label{eq:cmw_coupling}\\
 K&=\ca \left(\frac{67}{18}-\zeta_2\right)
 -\frac{10}{9}\tr n_f.
 \label{eq:cmw_K}
\end{align}    
\end{subequations}
Thus the correlated splitting of a soft gluon can be included at NLL accuracy
by evaluating the one-emission kernel with $\mathcal{A}^{\mathrm{CMW}}_i$ at
the transverse-momentum scale $\mu^2=\kt^2$~\cite{Catani:1990rr}.

Beyond this order there is no unique observable-independent continuation of
the CMW prescription.  The fixed-$\kt$ coupling was obtained through
$\mathcal{O}(\as^3)$ in Ref.~\cite{Banfi:2018mcq}, while the fixed-$\kt$ and
fixed-$\mt$ definitions and their relation to the cusp anomalous dimension were
studied in Ref.~\cite{Catani:2019rvy}.  The general family in
Eq.~\eqref{eq:general_soft_coupling} was analysed in
Ref.~\cite{Catani:2023kpw}.  Different choices of $\mathcal{F}$ first separate
at $\mathcal{O}(\as^3)$ in four dimensions, through terms controlled by the
running of the coupling.  
In the next section we study the soft-gluon coupling in the case of massive quark emitters. In particular, because we are interested in the generalisation of the CMW scheme, we will work at fixed $\kt$, i.e.\ we choose $\mathcal{F}=1$.
This choice is motivated by the fact that, in the context of parton showers, one applies the physical scheme at the level of the emission probability of a single on-shell gluon, for which $\mt = \kt$.

\section{Soft radiation off massive quarks}\label{sec:massive_emitters}

We now consider a heavy-quark dipole $\mathcal{Q}\bar{\mathcal{Q}}$ with hard momenta $p_1$ and
$p_2$, masses $p_1^2=m_1^2$ and $p_2^2=m_2^2$, and total invariant mass
$Q^2=(p_1+p_2)^2$.
To discuss the kinematics, we work in the centre-of-mass frame of the dipole. It is useful to introduce two light-like vectors $n_+$ and $n_-$ along the
space direction of the heavy quark momenta, normalised according to 
$n_+^2=n_-^2=0$ and $n_+\!\cdot n_-=2$:
\begin{align}
    n^\mu_{+}= N_+(1,0,0,1), \quad   n^\mu_{-}= N_-(1,0,0,-1),
\end{align}
where the above condition fixes $N_+ N_-=1$, but it does not fully determine them. We fix this remaining freedom by choosing a normalisation that gives a particularly simple light-cone decomposition of the hard momenta and simplifies the scalar products entering the eikonal current:
\begin{equation}
    N_\pm=\frac{m_1 e^{\mp\varphi/2}+ m_2 e^{\pm\varphi/2}}{Q},
\end{equation}
where we have defined the cusp angle:
\begin{align} \label{eq:cusp-angle-def}
    \varphi = \tanh^{-1}v, \quad v = \sqrt{1- \frac{m_1^2 m_2^2}{(p_1 \cdot p_2)^2}}.
\end{align}
We decompose the total soft momentum in the light-cone coordinates:
\begin{align}
 k^\mu= k_+ \frac{n^\mu_+}{2}+ k_- \frac{n^\mu_-}{2}+ \kt^\mu , \quad k^\mu= (k_+, k_-, \kt),
 \label{eq:massive_sudakov}
\end{align}
where $k_{\pm}= k \cdot n_{\mp}$, and $k^2= k_+ k_- -\kt^2$.
We also introduce for later convenience
\begin{equation}
 \alpha_r=\frac{m_r^2(p_s\!\cdot k)}
 {(p_r\!\cdot k)(p_r\!\cdot p_s)},
 \quad
 \omega^2=\frac{2(p_r\!\cdot k)(p_s\!\cdot k)}
 {p_r\!\cdot p_s}, \quad (r,s) =(1,2), \; (2,1). 
 \label{eq:massive_invariants}
\end{equation}
Note that with massive hard partons $\mt^2= k_+k_- \neq \omega^2$.

We define the rapidity of the emission $y$ with respect to the two light-like directions and the momentum fraction $z$ as
\begin{equation}
 y=\frac{1}{2}\ln\frac{k_+}{k_-}, \quad z = \frac{k_+}{p_{1+}}, 
 \label{eq:massive_rapidity}
\end{equation}
where we have also decomposed the hard momenta $p_1$ and $p_2$ in the light-cone basis:
\begin{equation}
 p^\mu_i=(p_{i+}, p_{i-},0), 
 \qquad p_{1\pm}= m_1e^{\pm\frac{\varphi}{2}}, \quad p_{2\pm}= m_2 e^{\mp\frac{\varphi}{2}}.
 \label{eq:massive_hard_momenta}
\end{equation}
We have kept the masses $m_1$ and $m_2$ distinct so that our discussion can also be applied, modulo a modification of the colour factor, to the case of a heavy-quark and gluon dipole, i.e.\ $m_1=m$, $m_2=0$. However, some care has to be taken when performing the single massless limit. Let us work this out. 
The rapidity defined in Eq.~\eqref{eq:massive_rapidity} can be written as
\begin{equation}
    y= \hat{y}+ y_\varphi,
\end{equation}
where $\hat{y} =\frac{1}{2}\ln \frac{k_0+k_z}{k_0-k_z}$ and
\begin{equation}
  y_\varphi=\frac{1}{2}\ln \frac{m_1 e^{\varphi/2}+ m_2 e^{-\varphi/2}}{m_1 e^{-\varphi/2}+ m_2 e^{\varphi/2}}
  = \frac{\varphi}{2}+ \ln \frac{m_1+m_2 e^{-\varphi}}{Q}, 
\end{equation}
is the rapidity shift due to the unequal masses, which vanishes when $m_1=m_2$. When $m_2\to 0$, we have that $\varphi\to \infty$. Consequently, $y_\varphi \to \infty$ and the rapidity $y$ becomes ill-defined. The single massless limit must be taken at fixed standard rapidity $\hat{y}\simeq y-\frac{\varphi}{2}-\ln\frac{m_1}{Q}$.
On the other hand, the $m_2\to 0$ limit for the longitudinal momentum fraction is smooth
\begin{equation}
    z=\frac{m_1 e^{\varphi/2}+ m_2 e^{-\varphi/2}}{m_1 e^{\varphi/2}}\frac{k_0+k_z}{Q} \xrightarrow[m_2\to0]{}\frac{k_0+k_z}{Q} .
\end{equation}

As discussed in the massless case,
the definition of the soft physical
coupling involves a prescription for replacing the total momentum $k$ of the soft radiation
with a massless momentum $p_k$. Following the discussion in the previous section, 
we consider projections that preserve the transverse momentum,
with the remaining freedom specified by the longitudinal variable held
fixed. 
To compare different projections, we define the double-differential (correlated) soft-radiation density as
\begin{equation}
 \frac{\de^2 \mathcal{S}_\mathcal{Q}}{\de  \kt^2 \de \zeta}
 =\frac{1}{2}\int_0^\infty\de k^2\, J_\zeta(k)\,
 w_{\mathcal Q}\bigl(k;\epsilon\bigr)
 =\frac{\kt^2}{2}\int_0^1 \frac{\de t}{t^2}\, J_\zeta \bigl(k(t,\kt,\zeta)\bigr)\,
 w_{\mathcal Q}\bigl(k(t,\kt,\zeta);\epsilon\bigr),
 \label{eq:massive_radiation_density_definition}
\end{equation}
where the dimensionless variable $t$ was introduced in Eq.~\eqref{eq:t_def} and $\zeta$ is the longitudinal variable that is held fixed in the projection. 
The Jacobian $J_\zeta(k)$ is defined by the $d$-dimensional integration measure, after integrating over the transverse angles:
\begin{equation}
   \de^d k
= \frac{1}{2}\,\mathcal{N}_\epsilon(\kt)\,
J_\zeta(k^2,\kt,\zeta)\,
\de k^2\, \de \kt^2\,\de \zeta.
\end{equation}
%
%
%

For a massive dipole, the symmetry under longitudinal boosts of the massless case is absent, and the leading-order density acquires a non-trivial dependence on  the longitudinal variables. This is easy to understand in that, physically, the emission probability is suppressed along the direction of the heavy quarks.
Symmetry alone, therefore, does not determine a preferred projection.
A useful additional criterion to identify suitable projections is to require that the ratio of the leading-order densities evaluated with the off-shell momentum $k$
and with the on-shell projected momentum $p_k$ is independent of the angle of the radiation in the dipole frame.
This mimics the way the class of corrections encoded in the splitting of the gluon (i.e.~cuts of self-energy diagrams) would contribute to the soft-radiation
density. The absence of artificial angular dependence indicates that the corresponding density will be proportional to the leading-order one, and hence could
be interpreted as a redefinition of the coupling.
This is the case when the rapidity is preserved by the projection, but not in other cases, as it will be shown later in this section. 
The above criterion does not imply that the full massive-dipole density is proportional to the Born one. Other non-Abelian contributions, not of the self-energy type can, in general, produce a non-trivial angular modulation when integrated inclusively with the projection. While this was not the case for massless dipoles, due to the symmetry argument made above, we will see an example of this phenomenon in the massive case.
In the following we will explicitly consider two projections, in which $\zeta=y$ and $\zeta=\ln z$, respectively. In both cases $J_\zeta=1$.
Since our projection preserves the transverse component $\kt$ of the total momentum $k$, if we fix also its rapidity then the longitudinal momentum fraction of the projected on-shell gluon $p_k$ will be $z_k = \frac{\kt}{p_{1+}}e^y \neq z$. Conversely, if we fix the longitudinal momentum fraction of the on-shell gluon to be the same as $z$, its rapidity would be $y_k = \ln \frac{z p_{1+} }{\kt} \neq y$.
We will investigate these two choices in the remainder of this section.

\subsection{LO calculation}
\label{sec:massive_LO}

The tree-level massive eikonal current squared reads
\begin{align}
 \left|J_{\mathcal Q}(k)\right|^2_{\mathrm{LO}}
 &=4\pi\cf\, \as(\mu^2)(\mu^2)^\epsilon S_\epsilon^{-1}
\left[\frac{2p_1\!\cdot p_2}
 {(p_1\!\cdot k)(p_2\!\cdot k)}
 -\frac{m_1^2}{(p_1\!\cdot k)^2}
 -\frac{m_2^2}{(p_2\!\cdot k)^2}\right].
 \label{eq:massive_lo_current}
\end{align}
Thus, the LO massive web function is simply given by
\begin{equation}
 w_{\mathcal Q,\mathrm{LO}}(k;0)
 =
 \cf\,  \frac{\as(\mu^2)}{2\pi}\delta_+(k^2)
\left[\frac{2p_1\!\cdot p_2}
 {(p_1\!\cdot k)(p_2\!\cdot k)}
 -\frac{m_1^2}{(p_1\!\cdot k)^2}
 -\frac{m_2^2}{(p_2\!\cdot k)^2}\right]
 \label{eq:massive_lo_web}.
\end{equation}
In order to expose the dependence on transverse and longitudinal degrees of freedom, we can evaluate the above expression in the massive dipole centre-of-mass frame,
which leads to the double-differential soft function at LO:
\begin{equation}
 \frac{\de^2 \mathcal{S}_{\mathcal{Q},\text{LO}}}{\de  \kt^2 \de y}
 = \frac{\cf}{\kt^2} \frac{\as}{\pi} \mathcal{B}(y,\varphi),
 \label{eq:massive_radiation_density_LO}
\end{equation}
with 
\begin{align}
\label{eq:massive_radiation_pattern}
    \mathcal{B}(y,\varphi)= \frac{\sinh^2 \varphi}{\left(\cosh\varphi+ \cosh{2y}\right)^2}.
\end{align}
Note that $\mathcal B(y,\varphi)\to 1$ in the massless limit, uniformly in rapidity. 
Furthermore, at LO we have $k^2=0$, and so there is no distinction between off-shell and on-shell kinematics.
 Then, we can simply use the definition in Eq.~\eqref{eq:massive_rapidity} to change variable in Eq.~\eqref{eq:massive_radiation_density_LO} and  express the LO double-differential soft function in terms of the longitudinal momentum fraction $z$:
\begin{equation}
     \frac{\de^2 \mathcal{S}_{\mathcal{Q},\text{LO}}}{\de  \kt^2 \de \ln z}
 = \frac{\cf}{\kt^2} \frac{\as}{\pi} \mathcal{B}\left( \frac{\varphi}{2}+\ln \frac{z m_1}{\kt},\varphi\right).
\end{equation}
This unambiguously defines the massive CMW coupling at this order as
\begin{equation}
   \mathcal{A}^{\mathrm{CMW}}_\mathcal{Q}(\as)
 \equiv\cf\frac{\as}{\pi} +\ord{\as^2}.
\end{equation}
 
\subsection{NLO calculation}
\label{sec:massive_NLO}
We now move to the central part of this work, namely the calculation of the massive web function at NLO. Its structure is the
direct analogue of
Eq.~\eqref{eq:massless_nlo_web}:
\begin{align}  \label{eq:massive_nlo_web}
 w_{\mathcal Q,\mathrm{NLO}}(k;\epsilon)
 =\mathcal N_\epsilon(\kt)\Bigg\{&
 \frac{\delta_+(k^2)}{(2\pi)^{d-1}}
 \left|J_{\mathcal Q}(k)\right|^2_{\mathrm{NLO}}
\\
 &+\int[\de k_1][\de k_2]\,
 \delta^{(d)}(k-k_1-k_2)
 \left[\frac12 W_{\mathcal Q}^{gg}(k_1,k_2)
       +n_fW_{\mathcal Q}^{q\bar q}(k_1,k_2)\right]\Bigg\}. \nonumber
\end{align}
The virtual correction is given by the NLO contribution to the soft function. It was computed analytically in ~\cite{Czakon:2014oma} and its explicit expression is reported in Appendix~\ref{app:calculations}.

\subsubsection{The \texorpdfstring{$n_f$}{nf} contribution}
\label{sec:massive_nf}
As in the massless case, we begin our discussion with the $n_f$ contribution, which
has the same splitting tensor as Eq.~\eqref{eq:massless_qq_splitting_tensor},
but it is contracted with the massive conserved current.  The two-body
integration therefore gives 
\begin{align}\label{eq:massive_qq_integrated_current}
 \int[\de k_1][\de k_2]\, \delta^{(d)}(k-k_1-k_2)
 &  W_{\mathcal Q}^{q\bar q}(k_1,k_2)
=\cf \tr
 \left(\frac{\as(\mu^2)}{\pi}\right)^2
 \frac{\mu^{4\epsilon}}{(k^2)^{1+\epsilon}}c(2\epsilon)
 \nonumber\\
 & \frac{(1-\epsilon)^2\Gamma(1-\epsilon)\Gamma(1-2\epsilon)}
 {\pi^{1-\epsilon}\Gamma(4-2\epsilon)}  
 \left[\frac{2p_1\!\cdot p_2}
 {(p_1\!\cdot k)(p_2\!\cdot k)}
 -\frac{m_1^2}{(p_1\!\cdot k)^2}
 -\frac{m_2^2}{(p_2\!\cdot k)^2}\right].
\end{align}
In order to compute this contribution to the double-differential soft function we have to integrate the above expression over the invariant mass of the soft pair or, equivalently, over the dimensionless variable $t$, as in Eq.~\eqref{eq:massive_radiation_density_definition}. It is clear that the result of this integral does depend on the choice of the longitudinal variable we keep fixed. 
If we fix the rapidity $y$ in the dipole rest-frame,\footnote{This argument can be generalised to any function of ratio $\frac{k_+}{k_-}$, as discussed in Appendix~\ref{app:calculations}.} the function $\mathcal{B}$ (defined in Eq.~\eqref{eq:massive_radiation_pattern}) factorises from the $t$ integral.  Adding the $n_f\tr$ part of the virtual correction, expanding
the endpoint singularity in plus distributions, and taking
$\epsilon\to0$ gives
\begin{equation}
 \label{eq:massive_nf_fixed_y}
  \frac{\de^2 \mathcal{S}_{\mathcal{Q},n_f}}{\de \kt^2 \de y}
 =\frac{\cf}{\kt^2} n_f\tr\left(\frac{\as(\mu^2)}{\pi}\right)^2
 \left[\frac13\ln\frac{\kt^2}{\mu^2}-\frac{5}9\right] \mathcal B(y,\varphi).
\end{equation}
Thus, at fixed $y$, the natural scale choice for this contribution is $\mu^2=\kt^2$, exactly as in
the massless calculation; the mass dependence appears only through the
LO radiation pattern  $\mathcal{B}$.

If we instead keep the longitudinal momentum fraction $z$ fixed, then the $t$-dependence of the eikonal factor in the square brackets of Eq.~\eqref{eq:massive_qq_integrated_current} does not factorise. Consequently, the resulting integral over $t$ becomes much more intricate.
The integration over $t$ of real and virtual contributions gives:
 \begin{align} \label{eq:nf-z-full}
 \frac{\de^2 \mathcal{S}_{\mathcal{Q},n_f}}{\de  \kt^2 \de \ln z}
& = \cf n_f \tr
\left(\frac{\as(\mu^2)}{\pi}\right)^2 \Bigg\{ \frac{2 p_1 \cdot p_2}{(p_1 \cdot p_k)(p_2 \cdot p_k)}\Bigg[\frac{1}{12} \ln \frac{2 z p_1 \cdot p_k}{\mu^2}-\frac{5}{36} \nonumber \\
 &+ \frac{ p_1 \cdot p_k}{12 z m_1^2}\left(-1+\coth\varphi\right)\ln \frac{m_2 e^{-\varphi}(p_1 \cdot p_k)}{m_1(p_2 \cdot p_k) } \Bigg]\nonumber \\
 &-\frac{m_1^2}{(p_1 \cdot p_k)^2} \Bigg(\frac{1}{12} \ln \frac{2 z p_1 \cdot p_k}{\mu^2}-\frac{2}{9}\Bigg) 
\nonumber \\
 & - \frac{m_2^2}{(p_2 \cdot p_k)^2} \left( \frac{1}{12} \ln \frac{2 z m_1 e^{\varphi} (p_2 \cdot p_k)}{\mu^2 m_2}- \frac{2}{9}\right)\Bigg\},
 \end{align}
where:
\begin{align}
  p_1 \cdot p_k= \frac{\kt^2+ z^2 m_1^2}{2 z}, \quad p_2 \cdot p_k= \frac{e^{-\varphi}m_2}{2 zm_1}\left(\kt^2 +z^2m_1^2e^{2\varphi} \right) , \quad p_1 \cdot p_2= m_1 m_2 \cosh \varphi.
 \end{align}
It is already clear that with this choice of longitudinal variable, even in the simplest contribution proportional to $n_f$, the massive eikonal factor does not factorise.
This implies that the fixed-$z$ projection is not suitable for a definition of a strong-coupling scheme. 
%

In order to better understand the structure of the result we employ a further approximation, namely the quasi-collinear limit~\cite{Catani:2000ef,Catani:2002hc} with respect to one of the hard partons.  In this approximation the transverse momentum 
of the emitted radiation, $\kt$, and the mass $m_i$ of the heavy quark are assumed to be small compared to the 
hard scale of the process but their ratio is assumed to be of order one. Considering, without loss of generality, the quasi-collinear limit with respect to $p_1$, we can set $m_2=0$ and Eq.~\eqref{eq:nf-z-full} becomes at leading power in $\kt^2, m_1^2$: 
\begin{align} \label{eq:nf-z-qc}
  \frac{\de^2 \mathcal{S}_{\mathcal{Q},n_f}^{\mathrm{qc}1}}{\de  \kt^2 \de \ln z}
 &=\left(\frac{\as(\mu^2)}{2\pi}\right)^2 \cf n_f \tr\Bigg\{\left( \frac{2p_1 \cdot p_2}{(p_1 \cdot p_k)(p_2 \cdot p_k)}-\frac{m_1^2}{(p_1 \cdot p_k)^2}\right) \\&\Bigg[\frac{1}{3} \ln\frac{2 z p_1 \cdot p_k}{\mu^2}- \frac{5}{9}\Bigg] \nonumber 
     +\frac{1}{3} \frac{m_1^2}{(p_1 \cdot p_k)^2}\Bigg\}. 
\end{align}
Note that with this choice of kinematical variables, we are not able to obtain a complete factorisation, even if the quasi-collinear limit is employed. This observation might have implications for the extension of quasi-collinear factorisation at higher orders, which would be interesting to investigate further.
Eqs.~\eqref{eq:nf-z-full},~\eqref{eq:nf-z-qc} suggest, however, that the scale of the coupling within the fixed-$z$ projection should be set to
\begin{equation}
\mu^2=2z\,p_1\!\cdot p_k=\kt^2+z^2m_1^2,
\end{equation}
as previously identified in the literature, see e.g.~\cite{Cacciari:2001cw,Dhani:2024gtx,Dhani:2025fbk,Ghira:2026vzw}.

Note that Eqs.~\eqref{eq:massive_nf_fixed_y} and
\eqref{eq:nf-z-full} (or its quasi-collinear approximation Eq.~\eqref{eq:nf-z-qc}) agree in the massless limit, but are different
finite-mass definitions, despite the fact that in both calculations the same transverse momentum is kept fixed. 
As already pointed out, $y$ and $z$ refer to the kinematics of the off-shell gluon and therefore they are not related by the on-shell relation between rapidity and longitudinal  momentum fraction.
Thus, the choice between keeping $y$ or $z$ fixed when performing the invariant-mass integration should be considered as a prescription. We will come back to the consequences of this prescription on the logarithmic accuracy in Sect.~\ref{sec:massive_other_variables}.

\subsubsection{Double-gluon contribution at fixed rapidity}
\label{sec:massive_gg}
We next discuss the non-Abelian contributions to the web function at NLO. In this section we describe the general strategy and the results, while leaving the details of the calculation in Appendix~\ref{app:calculations}.
Because of the factorisation properties previously discussed, we first focus on the calculation of the double-differential soft function at fixed rapidity.

The correlated double-gluon
matrix element is written in terms of the massive two-soft function
$S_{rs}$,
\begin{align}
 W_{\mathcal Q}^{gg}(k_1,k_2)
 &=-\left(4\pi\as(\mu^2)\mu^{2\epsilon}S_\epsilon^{-1}\right)^2
 \ca\sum_{r,s=1}^{2}\mathbf T_r\!\cdot\mathbf T_s\,
 S_{rs}(k_1,k_2)
 \nonumber\\
 &=\left(4\pi\as(\mu^2)\mu^{2\epsilon}S_\epsilon^{-1}\right)^2
 \ca\cf\sum_{r\ne s}
 \left[S_{rs}(k_1,k_2)-S_{rr}(k_1,k_2)\right].
 \label{eq:massive_gg_Sij}
\end{align}
As we did in the massless case, we separate off the contribution that only depends on the total soft momentum $k_1+k_2$ from the remaining correlated contributions:
\begin{equation}  \label{eq:massive_Sij_decomposition}
 S_{rs}(k_1,k_2)
 =\widehat S_{rs}^{m=0}(k_1,k_2)
 +\widetilde S_{rs}^{m=0}(k_1,k_2)
 +m_r^2\widetilde S_{rs}^{m\ne0}(k_1,k_2)
 +m_s^2\widetilde S_{sr}^{m\ne0}(k_1,k_2).
\end{equation}
The label $m=0$ identifies the structures that survive in the massless limit; they have the same functional form as in the massless case but they retain their dependence on $m_1$ and $m_2$ through the scalar products of the four momenta.  The $m\neq0$ label instead identifies new structures that appear in the massive case~\cite{Czakon:2014oma}. Explicit expressions are given in Appendix~\ref{app:masters}.

As in the massless case, the $\widehat{S}$-part can be obtained immediately from
the $n_f$ contribution by the replacement~\eqref{eq:massless_gg_replacement}. Hence
Eq.~\eqref{eq:massive_qq_integrated_current}, together with~\eqref{eq:massless_gg_replacement}, is an explicit result for the
$\widehat S$-contribution.  This term contains the real-emission term needed to cancel the remaining part of the $\beta_0/\epsilon$ pole of the virtual correction.
The sum of real and virtual contributions leads to
\begin{align}
\label{eq:soft-function-S-hat}
    \frac{\de^2 \widehat{\mathcal{S}}_{\mathcal{Q}}}{\de \kt^2 \de y}=& \left(\frac{\as(\mu^2)}{\pi}\right)^2 \frac{\cf \ca}{\kt^2}  \left(\frac{11}{12}  \ln \frac{\mu^2}{\kt^2} + \frac{67}{36}  \right)\mathcal{B}(y,\varphi).
\end{align}

The integration of the remaining terms, namely $\widetilde S_{rs}^{m=0}$ and $\widetilde S_{rs}^{m\neq0}$, is much more intricate. We follow the strategy of Ref.~\cite{Catani:2023tby} , writing the angular integrals in terms of the masters \cite{Somogyi:2011ir}:
\begin{align}
 \mathcal I_{ab}^{\pm}(a_r,a_s,\chi;\epsilon)
 =\int_0^\pi\!\de\theta\int_0^\pi\!\de\phi\,
 \frac{(\sin\theta)^{1-2\epsilon}(\sin\phi)^{-2\epsilon}}
 {(1-a_r\cos\theta)^a
  (1\pm a_s\cos\chi\cos\theta
     \pm a_s\sin\chi\sin\theta\cos\phi)^b},
 \label{eq:massive_angular_masters}
\end{align}
where:
    \begin{equation}
 a_r^2=1-\frac{k^2m_r^2}{(p_r\!\cdot k)^2}
      =1-\frac{2\alpha_r k^2}{\omega^2},
 \qquad
 a_ra_s\cos\chi
 =1-\frac{k^2p_r\!\cdot p_s}
 {(p_r\!\cdot k)(p_s\!\cdot k)}
 =1-\frac{2k^2}{\omega^2}.
 \label{eq:massive_two_body_geometry}
\end{equation}
The advantage of the rapidity prescription becomes evident in this case.  At fixed $y$ the
quantities $v$, $\alpha_1$, and $\alpha_2$ are independent of $t$, while
$t\omega^2$ is constant.  The invariant-mass integration can be performed analytically. After expansion in plus distributions, all
$1/\epsilon^2$ and $1/\epsilon$ poles cancel between the real-emission contributions and the virtual corrections. Details are given in Appendix~\ref{app:calculations}.
The result for the double-differential soft function can be written as
\begin{align} 
  \frac{\de^2 \mathcal{S}_{\mathcal{Q}}}{\de  \kt^2 \de y}
 &=\frac{\cf}{\kt^2}\frac{\as(\mu^2)}{\pi}\mathcal B(y,\varphi)\left[1+\frac{\as(\mu^2)}{2\pi}\left(-2 \pi \beta_0 \ln \frac{\kt^2}{\mu^2}+  \frac{67}{18}\ca-\frac{10}{9} n_f \tr \right)\right]
 \nonumber\\
 &\quad +\left(\frac{\as(\mu^2)}{\pi}\right)^2 
 \frac{\cf\ca}{2 \kt^2}\,F\!\left(y,\varphi\right).
 \label{eq:massive_fixed_y_result}
\end{align}
The term in brackets proportional to the LO result uniquely stems from the $\widehat{S}$ and $n_f$ real-emission corrections combined with the $\beta_0$ contribution from the virtuals. These terms are sufficient to confirm $\mu^2=\kt^2$ as the natural choice for the scale of the running coupling. They also reproduce the rational part of the CMW coefficient $K$. 
The remaining non-Abelian function contains the non-$\beta_0$ part of the one-loop current together with the
diagonal, massless-like, and genuinely massive $\widetilde S$ contributions.
We find: 
\begin{align} \label{eq:full_F}
F(y,\varphi)
&=\frac{1}{c_{\varphi,y}^{\,2}}\Bigg\{(\cosh \varphi \, c_{\varphi,y}-\sinh^2 \varphi)
 \bigl[\coth\varphi \;\mathcal C(y,\varphi)-\varphi^2\bigr]
+\sinh^2 \varphi\bigl(6\zeta_2-4y^2\bigr)
\nonumber\\
&+\cosh \varphi(\varphi \sinh \varphi-c_{\varphi,y})
 \ln(2c_{\varphi,y})
+\varphi \cosh \varphi\sinh(2y)\,\mathcal L(y,\varphi)\nonumber\\
&
+\sinh \varphi \cosh \varphi
 \left[
 2\varphi-\varphi^2-6\zeta_2
 -\operatorname{Li}_2(1-e^{-2\varphi})
 \right]
\nonumber\\
&+(\cosh \varphi \,c_{\varphi,y}-2\sinh^2 \varphi)
 \left[
 \frac{\varphi \sinh \varphi+2y\sinh(2y)}{c_{\varphi,y}}
 -2y\coth\varphi\,\mathcal L(y,\varphi)
 \right]\Bigg\},
\end{align}
where
\begin{align}
& c_{\varphi,y} = \cosh \varphi+\cosh(2y),
\quad
\mathcal L(y,\varphi)
= \ln\frac{\cosh(y+\varphi/2)}
              {\cosh(y-\varphi/2)},
\nonumber \\
&\mathcal C(y,\varphi)
= \operatorname{Li}_2(-e^{-\varphi-2y})
 + \operatorname{Li}_2(-e^{-\varphi+2y})
 + \frac{\varphi^2}{2}+2y^2+\zeta_2.
\end{align}
The massless
limit of Eq.~\eqref{eq:full_F} provides a stringent check:
\begin{equation}
 \lim_{\varphi\to \infty}F\left(y,\varphi\right)=-\zeta_2 ,
 \label{eq:massive_FA_massless_limit}
\end{equation}
so Eq.~\eqref{eq:massive_fixed_y_result} reduces to the CMW result.

\subsubsection{Comments on the momentum-fraction prescription}
\label{sec:massive_other_variables}

For completeness, we also report the structure of the result at fixed longitudinal momentum fraction. In this case, we are able to compute explicitly all $\beta_0$ contributions, while we have not performed the calculation of the correlated part. We find
\begin{align} 
  \frac{\de^2 \mathcal{S}_{\mathcal{Q}}}{\de  \kt^2 \de \ln z}
 &=\frac{\cf}{\kt^2}\frac{\as(\mu^2)}{\pi} \mathcal{B}\left( \frac{\varphi}{2}+\ln \frac{z m_1}{\kt},\varphi\right)\left[1+\frac{\as(\mu^2)}{2\pi}\left(-2 \pi \beta_0 \ln \frac{\kt^2+z^2 m_1^2}{\mu^2} \right. \right.\nonumber\\
 &\left. \left.+  \frac{67}{18}\ca-\frac{10}{9} n_f \tr \right)\right]
  +\left(\frac{\as(\mu^2)}{\pi}\right)^2 
 \frac{\cf}{\kt^2}\,G\!\left(z, \varphi,
 \frac{m_1^2}{\kt^2},\frac{m_2^2}{\kt^2}\right),
 \label{eq:massive_fixed_z_result}
\end{align}
where we have grouped in the function $G$ all the non-factorising contributions.

Although we are not going to define the soft-gluon coupling for this prescription, it is interesting to inspect whether the choice of scale $\mu^2=\kt^2 +z^2 m_1^2$ leads to a different logarithmic structure than $\mu^2=\kt^2$. In particular, we study the rather general case of (jet) shape variables. Thus, we consider an observable that in the soft and collinear limit is parametrised as $\lambda_\alpha= \frac{\kt^\alpha}{Q^\alpha} z^{1-\alpha}$. In the quasi-collinear limit, the leading contribution to the differential distribution is given by
\begin{equation}
    \frac{1}{\sigma_0} \frac{\de \sigma}{\de \lambda_\alpha}= \int_0^1 \frac{\de z}{z}\int_0^{Q^2} \frac{\de \kt^2 \; \kt^2}{(\kt^2+z^2 m_1^2)^2}\frac{\cf \as (\mu^2)}{\pi}
    \, \delta \left(\left(\frac{\kt}{Q}\right)^\alpha z^{1-\alpha}-\lambda_\alpha\right).
\end{equation}
We compute the difference between the distributions obtained with the two scale choices $\mu^2=\kt^2$ or  $\mu^2=\kt^2+z^2 m_1^2$. 
For simplicity, we focus on the case $\alpha\ge 1$ (but we have checked that the conclusion holds also for the general case $\alpha>0$). We obtain
\begin{align}
    \Delta \sigma&= -\frac{\as^2 \cf \beta_0}{\pi}\int_0^1 \frac{\de z}{z} \int_0^1 \frac{u \de u}{(u+z^2 \frac{m_1^2}{Q^2})^2}\ln \frac{u}{u+z^2 \frac{m_1^2}{Q^2}}  \delta \left(u^\frac{\alpha}{2}z^{1-\alpha} -\lambda_\alpha \right) +\order{\as^3}\nonumber\\
    &
= \frac{\as^2 \cf \beta_0 (\zeta_2-1)}{\pi \lambda_\alpha} +\dots.
\end{align}
where the dots stand for subleading contributions at small $\lambda_\alpha$.
Thus, the difference between the two prescriptions is an NNLL effect, as already noted in Ref.~\cite{Ghira:2023bxr}.

\section{Soft-coupling schemes with massive quarks}\label{sec:pheno}

In this section we exploit the results obtained with the rapidity prescription to define the effective coupling in the presence of massive emitters. Before doing so, we link our findings to existing calculations in the literature. 
Not only this allows us to perform quantitative checks of our calculations, but also offers insights in the definition of the effective coupling. 

\begin{figure}
    \centering
\includegraphics[width=1.0\linewidth]{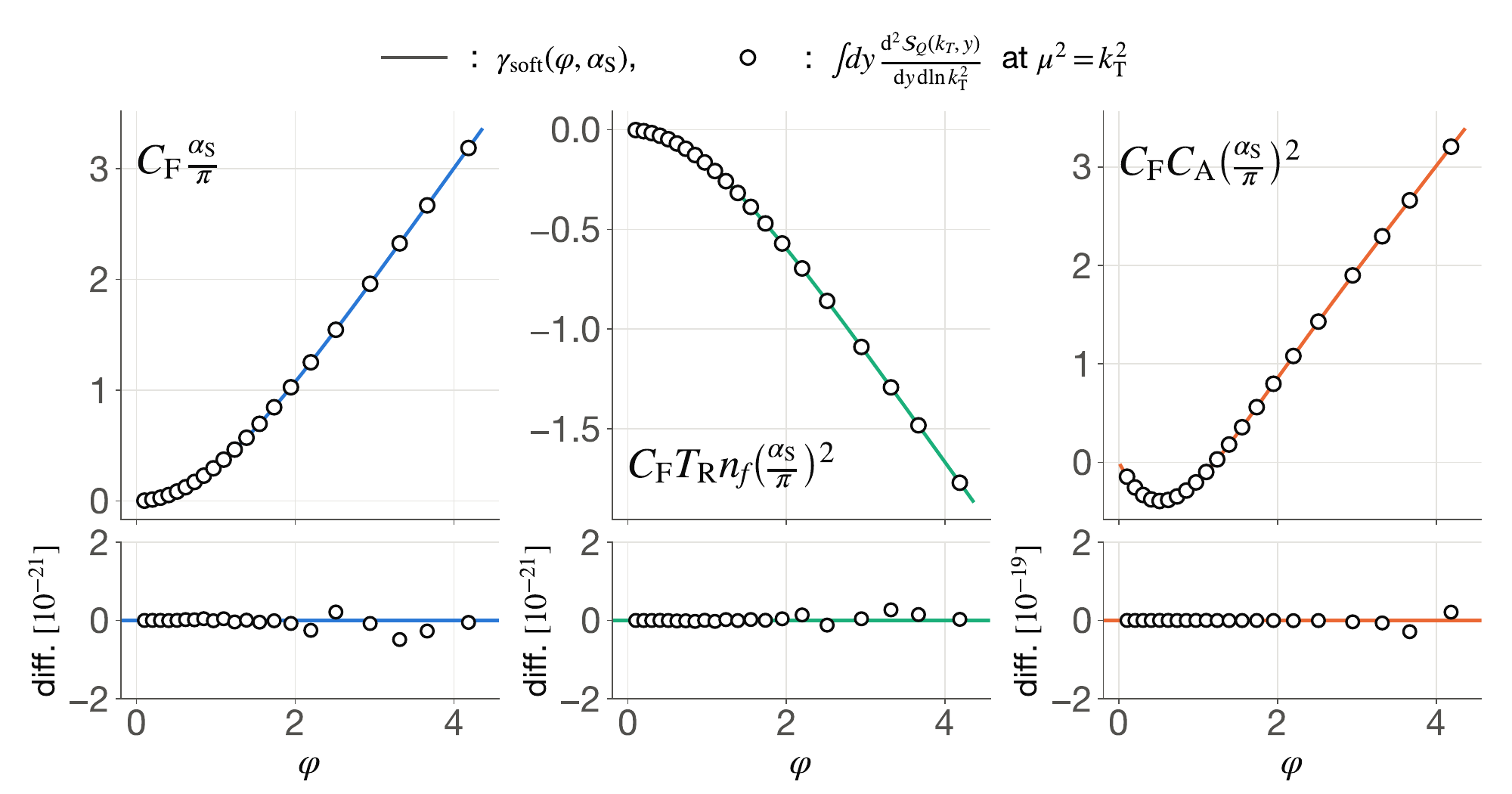}
    \caption{Numerical check of Eq.~\eqref{eq:intS-gamma-soft}. We check our result for each colour structure, namely $\cf$, $\cf \tr n_f$, and $\cf \ca$.}
    \label{fig:anomalous-dimen-checks}
\end{figure}

The soft function that we have computed is intimately related to the massive soft anomalous dimension or, equivalently, for two Wilson lines, the massive cusp anomalous dimension~\cite{Korchemsky:1987wg,Kidonakis:2009ev,Becher:2009kw,vonManteuffel:2014mva}. Namely, if we set $\mu^2=\kt^2$, we have 
\begin{align}\label{eq:intS-gamma-soft}
     \int_{-\infty}^{+\infty} \de y\, \kt^2\frac{\de^2 \mathcal{S}_{\mathcal{Q}}}{\de  \kt^2 \de y}
=        \gamma_\text{soft}(\varphi, \as),
\end{align}
where
\begin{equation}\label{eq:gamma-soft-expansion}
    \gamma_\text{soft}(\varphi, \as)=
    \cf \frac{\as}{\pi}
    \left[\gszero(\varphi)
    +\frac{\as}{2 \pi }\left( K \gszero(\varphi) + \ca \gsone(\varphi)
    \right) \right],
\end{equation}
with 
\begin{subequations}
\begin{align}
    \gszero(\varphi)&=\varphi \coth \varphi -1, \\
    \gsone(\varphi)&=1-(\coth\varphi-1) \left[5 \zeta_2(\varphi\coth\varphi-1)-
    \frac{\varphi^2}{3}(\varphi\coth\varphi-3)\right] 
    \\
&-\coth\varphi \left[ 2\varphi\,\ln\!\big(1-e^{-2\varphi}\big)-(1+\varphi\coth\varphi)\,\mathrm{Li}_2\big(e^{-2\varphi}\big)\right]+\coth^2\!\varphi\,\Big[\mathrm{Li}_3\big(e^{-2\varphi}\big)-\zeta_3\Big]. \nonumber
\end{align}
    \end{subequations}
We note that the above expression can be easily compared to the result of Ref.~\cite{Becher:2009kw}, which is already written in terms of the cusp angle. 
However, a few remarks are in order. First, the soft anomalous dimension $\gamma_\text{soft}(\varphi, \as)$ corresponds to the full anomalous dimension of Ref.~\cite{Becher:2009kw}, namely the sum of their massive cusp term ($\gamma_\text{cusp}$) and the quark anomalous dimension ($\gamma^Q$). Second, their result is written for spacelike kinematics, while ours is for the timelike case. Thus, in order to compare the two, we have to analytically continue the result of Ref.~\cite{Becher:2009kw} to timelike kinematics: $\varphi \to \varphi - i \pi$. After this replacement, the anomalous dimension of Ref.~\cite{Becher:2009kw} develops an imaginary part and $\gamma_\text{soft}$ is its real part:~\footnote{We thank HuaXing Zhu for discussions about this point.}
\begin{equation}
    \gs(\varphi,\as)=\text{Re}\left[\cf\gamma_\text{cusp}(\varphi-i\pi,\as)+ 2 \gamma^Q(\as) \right].
\end{equation}
Thus, as a check of our calculation,  we numerically integrate our soft function over rapidity and compare it to the massive soft anomalous dimension, Eq~\eqref{eq:gamma-soft-expansion}. 
The results of this comparison is plotted in Figure~\ref{fig:anomalous-dimen-checks} and they show excellent agreement. 

Let us now discuss possible definitions of the effective coupling.
In order to define a soft-coupling scheme, we have to factor a universal contribution out of the eikonal term $\mathcal{B}$. This may depend on the masses and $Q^2$, but should not depend on any other kinematical variable. We have seen that the rational part of $K$ factors out naturally if we perform the calculation at fixed rapidity. 
While the result of the calculation is given in Eq.~\eqref{eq:massive_fixed_y_result}, one can reorganise it in different ways as the sum of a term that multiplies the leading-order kernel $\mathcal B(y,\varphi)$ and a non-factorising remainder that contributes at the same perturbative order.
Thus, the most straightforward option is to also factor the remaining non-rational part. This amounts to separating the non-Abelian remainder as
\begin{equation}
    F(y,\varphi)= -\zeta_2 \mathcal{B}(y,\varphi)+R_1(y,\varphi),
\end{equation}
so that we have
\begin{align}
    \kt^2 \frac{\de^2 \mathcal{S}_{\mathcal{Q}}}{\de  \kt^2 \de y}&= \cf\frac{\as}{\pi}
     \left[\left(1+\frac{\as}{2\pi} K \right)\mathcal B(y,\varphi)
  +\frac{\as}{2\pi}
 \ca R_1\!\left(y,\varphi\right)\right],
\end{align}
where 
\begin{equation}
      \int_{-\infty}^{+\infty} \de y\, R_1(y,\varphi)=\gsone(\varphi),
\end{equation}
and we can define the CMW coupling to be the same as the massless one, Eq.~\eqref{eq:cmw-definition}, with, in addition, a remainder proportional to $\ca$.

An alternative way to organise Eq.~\eqref{eq:massive_fixed_y_result} and define a physical scheme is to separate $F$ as
\begin{equation}
    F(y,\varphi) = \left(-\zeta_2 + \frac{\gsone(\varphi)}{\gszero(\varphi)}\right)\mathcal{B}(y,\varphi)+R_2(y,\varphi),
\end{equation}
so that we have
\begin{align}
    \kt^2 \frac{\de^2 \mathcal{S}_{\mathcal{Q}}}{\de  \kt^2 \de y}&= \cf\frac{\as}{\pi}
     \left[\left(1+\frac{\as}{2\pi}\left( K+ \ca \frac{\gsone(\varphi)}{\gszero(\varphi)}\right)\right)\mathcal B(y,\varphi)
  +\frac{\as}{2\pi}
 \ca R_2\!\left(y,\varphi\right)\right],
\end{align}
where the function $R_2$ integrates to zero:
\begin{equation}
      \int_{-\infty}^{+\infty} \de y\, R_2(y,\varphi)=0.
\end{equation}
Thus, we can define the massive CMW coupling as 
\begin{subequations}\label{eq:massive_CMW_us}   
\begin{align}
\mathcal{A}^{\mathrm{CMW}}_\mathcal{Q}(\as)
 &\equiv \cf\frac{\as}{\pi}\left(1+ \frac{\as}{2 \pi } K_\mathcal{Q}\right)  +\ord{\as^3}, \\ K_\mathcal{Q}&=K+\ca \frac{\gsone(\varphi)}{\gszero(\varphi)}. 
\end{align}
\end{subequations}
In the massless limit, $\varphi \to \infty$, we recover the standard CMW coupling $K_\mathcal{Q}\to K$. In the opposite (threshold) limit the cusp angle $\varphi$ and, hence the velocity $v$, goes to zero and we have  $K_\mathcal{{Q}} \to -\ca \frac{\pi^2}{v} $. Therefore, this soft coupling displays a Coulomb enhancement at small velocities.

\begin{figure}
    \centering
     \includegraphics[width=0.85\linewidth]{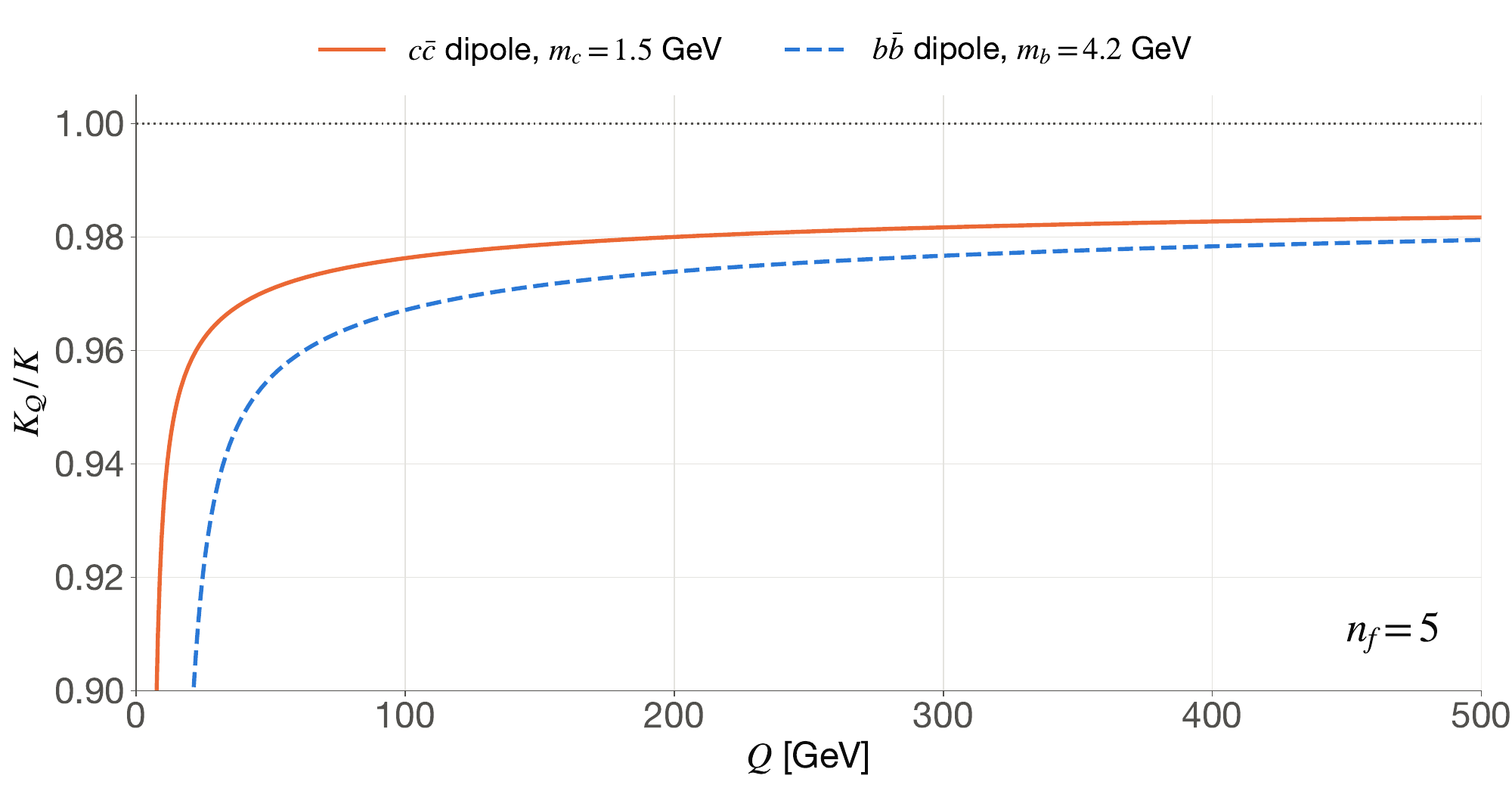}
    \caption{Plots of the ratio $K_\mathcal{Q}/K$ for a $c \bar c$ dipole (in solid orange) and for a $b \bar b$ dipole (in dashed blue), as a function of the dipole invariant mass $Q$, for fixed $n_f=5$.}
    \label{fig:KQ}
\end{figure}
It is interesting to quantitatively study the numerical size of the scheme coefficient $K_\mathcal{Q}$. To this end, using the definition of the cusp angle, Eq~\eqref{eq:cusp-angle-def}, we plot in Figure~\ref{fig:KQ} the ratio $K_{\mathcal{Q}}/K$, as a function of the dipole invariant mass $Q$, for the case of a charm ($c$) quark, i.e.\ $m_1=m_2=m_c$, and for a bottom ($b$) quark, i.e.\ $m_1=m_2=m_b$, for fixed $n_f=5$.  At large $Q$, mass effects becomes smaller and the ratio approaches one as an inverse power of the mass logarithm. At the opposite end, at small-$Q$, the effects are larger because of the Coulomb behaviour previously discussed. 
However, the difference between the two schemes is below 5\% for a wide range of~$Q$.

The two variants of the schemes discussed above have both advantages and shortcomings. The choice with $K$, i.e.\ the same as in the massless case, is simple and it follows almost directly from the calculation we have presented. It puts on firmer ground what is already employed in many phenomenological applications. On the other hand, the use $K_\mathcal{Q}$ allows us to include in the coupling the full soft anomalous dimension, but it introduces the $1/\gszero$ behaviour that only cancels when multiplied by the eikonal contribution $\mathcal{B}$.

A direct application of our results is heavy-flavour jet substructure, as highlighted in the introduction to this paper. 
In those calculations, heavy-quark mass effects were incorporated in the one-emission radiator through the massive quasi-collinear splitting function, while the inclusive non-Abelian two-loop correction was described using the ordinary massless CMW coupling. The inclusion of $K_{\mathcal Q}$  would not modify the NLL accuracy of those predictions.
Indeed, we have shown that different choices of the scale of the coupling discussed in this work lead to differences that are NNLL. 
 Nevertheless, we believe that our analysis provides useful information for extending this resummation framework to NNLL accuracy. Since $K_{\mathcal Q}$ depends only on the cusp angle of the underlying massive dipole, it can be incorporated into the Sudakov radiator without modifying the observable-dependent measurement function or introducing additional dependence on the resolved-emission kinematics. The result thus identifies a universal, observable-independent ingredient for future NNLL studies of heavy-flavour jet substructure and provides a non-trivial consistency condition on their two-loop soft contribution. 

Our results also have a natural application to the development of higher-accuracy parton showers with massive quarks, discussed in the introduction. Parton showers commonly incorporate inclusive two-loop soft effects through the massless CMW coupling, including for branchings involving massive partons. The coefficient $K_{\mathcal Q}$ provides a systematically replacement for a massive colour dipole. Importantly, it depends only on the cusp angle between the massive eikonal lines (and hence on the heavy-quark velocities), and not on the kinematics of the resolved soft emission. It can therefore be incorporated as a dipole-dependent correction to the effective soft coupling, while leaving the one-emission eikonal kernel and its phase-space dependence unchanged. This prescription reproduces the correct inclusive two-loop soft radiation and interpolates smoothly between finite-mass configurations and the massless limit. It consequently represents an important ingredient for parton showers aiming at controlled logarithmic accuracy for heavy-flavour observables.

\section{Conclusions and Outlook}\label{sec:conclusions}

In this work, we have addressed the definition of the soft-gluon effective coupling in QCD in the presence of massive hard emitters. We have considered a dipole formed by two particles with generic masses $m_1$ and $m_2$, and computed the corresponding double-differential soft function through $\ord{\as^2}$. In particular, we investigated the distributions differential in the transverse momentum $\kt$ of the radiation and in either its rapidity $y$ or its longitudinal momentum fraction $z$.

The two choices of longitudinal variables ($y$ and $z$) lead to different results for the inclusive integration over the invariant mass of the unresolved system. At fixed rapidity, the invariant-mass integration can be performed analytically and exhibits particularly transparent factorisation properties. 
Our calculation combines the one-loop massive soft current with the correlated emission of two soft gluons or a light quark-antiquark pair. We have checked the final result against the known two-loop soft anomalous dimension for massive quarks. 
By contrast, keeping the longitudinal momentum fraction fixed gives rise to more involved, non-factorising structures.
In this case, we have computed explicitly the $\beta_0$ contribution and discussed the general structure of the remaining non-Abelian part.

For the rapidity prescription, our complete result allows us to formulate soft-coupling schemes for massive emitters. Unlike in the massless case, we have found that, already at $\ord{\as^2}$, there is some freedom in deciding which finite massive contributions should be absorbed into the effective coupling and which should be retained in a non-Abelian remainder function. We have investigated two natural choices. In the first, the coupling is defined using the same CMW coefficient as for massless emitters, while all additional mass-dependent terms are assigned to the remainder. In the second, the coupling absorbs the complete massive soft anomalous dimension. The corresponding remainder then integrates to zero over rapidity and describes only a redistribution of radiation within the rapidity profile.
Numerically, the difference between the resulting coefficients $K$ and $K_{\mathcal Q}$ is at the level of a few percent for charm- and bottom-quark dipoles with invariant masses ranging from a few tens to a few hundreds of GeV. The difference becomes more pronounced close to the heavy-quark production threshold, where Coulomb-enhanced contributions to the massive anomalous dimension become important.

We find that the momentum-fraction prescription is instead unsuitable for defining an effective coupling scheme, as the massive eikonal does not factorise.
Nevertheless, it is instructive to study the terms proportional to the QCD beta function to determine the natural scale of the running coupling. In the rapidity prescription, this scale is $\mu^2=\kt^2$, as in the massless case. In the longitudinal-momentum-fraction prescription, the corresponding scale is instead $\mu^2=\kt^2+z^2m^2$.
We have verified that the difference between these prescriptions affects standard shape observables only starting at NNLL accuracy, consistently with previous analyses of heavy-quark fragmentation functions.

Our calculation highlights several subtleties in the description of soft radiation from heavy quarks. Although the usual CMW coupling remains sufficient to achieve NLL accuracy for standard observables, massive kinematics introduces prescription and scheme dependency whose effects first become relevant at NNLL. A systematic treatment of these effects will therefore be required for NNLL resummations of heavy-flavour jet observables.
We have not attempted such a resummation here. Nevertheless, the soft functions and effective-coupling schemes derived in this work provide the ingredients needed for this future development. 
Furthermore, they should also be useful in constructing higher-accuracy parton showers that consistently incorporate finite-mass effects and correlated soft radiation. We leave the exploration of these applications to future work.

\paragraph{Acknowledgments.}
The work of AG is supported by the Excellence Cluster ORIGINS, funded
by the Deutsche Forschungsgemeinschaft (DFG, German Research Foundation) under Germany’s Excellence Strategy — EXC-2094-390783311 and by a TUM Global Postdoc Fellowship.
The work of SD received support from the FWO (contract No. 1227426N).
AG and SM acknowledge support from the CERN Theoretical Physics Department under the CERN-TH Visitor Programme.
The work of PM is funded by the European Union (ERC, grant agreement No. 101044599). Views and opinions expressed are however those of the authors only and do not necessarily reflect those of the European Union or the European Research Council Executive Agency. Neither the European Union nor the granting authority can be held responsible for them.

\appendix
\section{Details of the calculation:  $\epsilon$-expansion of the angular integrals}\label{app:masters}
In this section we report the expansion in the dimensional regulator $\epsilon$ of the angular master integrals, following \cite{Catani:2023kpw}.
As stated in the main text in Eq.~\eqref{eq:massive_Sij_decomposition}, we decompose the result into four distinct contributions. First, the $\widehat{S}$ contributions can be computed analogously to the massless case, despite the need to contract the projector with two massive conserved currents.
Moreover, we have that the contributions $S^{m=0}_{ij}(k_1,k_2), S^{m\neq0}_{ij}(k_1,k_2)$ read~\cite{Czakon:2014oma, Catani:2023kpw}:
\begin{subequations}
\begin{align}
    \tilde{S}^{m=0}_{rs}(k_1,k_2)&= - \frac{(p_r\!\cdot\! p_s)^2}{2(p_r\!\cdot\! k)(p_s\!\cdot\! k)}
\Bigg[
\frac{2}{(p_r\!\cdot\! k_1)(p_s\!\cdot\! k_1)}
+ \frac{1}{(p_r\!\cdot\! k_1)(p_s\!\cdot\! k_2)}\Bigg]+ \frac{2 p_r \cdot p_s}{k^2(p_r\cdot k_1)(p_s\cdot k_2)}
\nonumber\\
&\quad
- \frac{(p_r\!\cdot\! p_s)}{2k^2(p_r\!\cdot\! k)(p_s\!\cdot\! k)}
\frac{\big[(p_r\!\cdot\! k_1)(p_s\!\cdot\! k_2)
- (p_r\!\cdot\! k_2)(p_s\!\cdot\! k_1)\big]^2}
{(p_r\!\cdot\! k_1)(p_s\!\cdot\! k_2)
(p_r\!\cdot\! k_2)(p_s\!\cdot\! k_1)}
+ (k_1 \leftrightarrow k_2), \\
 \tilde S^{m\neq 0}_{rs}(k_1,k_2)&=\frac{(p_r\cdot p_s)}{2(p_r\cdot k)^2}\Bigg[ \frac{1}{(p_r\cdot k_1)(p_s\cdot k_1)}+\frac{1}{(p_r\cdot k_1)(p_s\cdot k_2)}+\frac{1}{(p_r\cdot k_2)(p_s\cdot k_1)}+\frac{1}{(p_r\cdot k_2)(p_s\cdot k_2)}\Bigg] \nonumber \\
    &-\frac{1}{k^2(p_r\cdot k)^2}\Bigg[ \left(p_s\cdot k\right)(p_r\cdot k) \left(\frac{1}{(p_s\cdot k_1)(p_r\cdot k_2)}+\frac{1}{(p_s\cdot k_2)(p_r\cdot k_1)}\right)\nonumber \\
    &- 3(p_r\cdot k) \left(\frac{1}{(p_r\cdot k_1)}+\frac{1}{(p_r\cdot k_2)}\right)+\frac{(p_r\cdot k)}{(p_s\cdot k)}\left(\frac{(p_s\cdot k_2)}{(p_r\cdot k_1)}+\frac{(p_s\cdot k_1)}{(p_r\cdot k_2)}\right)+2\Bigg].
\end{align}
\end{subequations}
Finally, we note that for $r=s$, we have:
\begin{align}
    \int[\de k_1][\de k_2]\delta^d(k-k_1-k_2)&\left(\tilde{S}^{m=0}_{rr}(k_1,k_2)+2 m_r^2\tilde{S}^{m\neq0}_{rr}(k_1,k_2)\right)\nonumber  \\
    &  =  \int[\de k_1][\de k_2]\delta^d(k-k_1-k_2) \tilde{S}_{rr}(k_1,k_2),
\end{align}
where:
\begin{align}
    \tilde{S}_{rr}(k_1,k_2)= \frac{2m_r^2}{(p_r \cdot k_1)(p_r\cdot k) k^2} \left(\frac{m_r^2 k^2}{(p_r\cdot k)^2}+4\right).
\end{align}
From the above expression we can derive:
\begin{subequations}
\begin{align}
    f_{rr}(a_r; \epsilon)=& \left(5- a_r^2\right) \mathcal{I}_{10}(a_r; \epsilon), \\
    f_{rs}^{m=0}(a_r,a_s,\chi; \epsilon)=& \mathcal{I}^+_{11}(a_r, a_s, \chi;\epsilon) (1+ a_r a_s \cos \chi)-2\mathcal{I}^-_{11}(a_r, a_s, \chi;\epsilon) (1-a_r a_s \cos \chi)\nonumber \\
    +& \mathcal{I}_{10}(a_r;\epsilon)+ \mathcal{I}_{01}(a_s;\epsilon), \\
    f_{rs}^{m\neq0}(a_r,a_s,\chi; \epsilon)=& -\mathcal{I}^+_{11}(a_r, a_s, \chi;\epsilon) (1+ a_r a_s \cos \chi)+\mathcal{I}^-_{11}(a_r, a_s, \chi;\epsilon) (1-a_r a_s \cos \chi)\nonumber \\
    -& \frac{1}{2}\mathcal{I}^{+}_{1-1}(a_r, a_s, \chi;\epsilon)+ 3 \mathcal{I}_{10}(a_r;\epsilon)- \frac{1}{2}\mathcal{I}_{00}.
\end{align}
\end{subequations}
Since we are interested in the calculation of the double differential soft function, we can perform a further simplification. Following the strategy of \cite{Catani:2023kpw}, we divide each contributions as: 
\begin{align}
 f_{rr}(a_r; \epsilon)=& -\frac{\pi}{\epsilon}\left(\sigma_{rr}(a_r; \epsilon)+ \epsilon\rho_{rr}(a_r; \epsilon)\right) , \\
    f_{rs}^{m=0}(a_r,a_s,\chi; \epsilon)=&-\frac{\pi}{\epsilon}\left(\sigma^{m=0}_{rs}(a_r, a_s, \chi; \epsilon)+ \epsilon\rho^{m=0}_{rs}(a_r, a_s, \chi; \epsilon)\right) , \\
    f_{rs}^{m\neq0}(a_r,a_s,\chi; \epsilon)=&-\frac{\pi}{\epsilon}\left(\sigma^{m\neq0}_{rs}(a_r, a_s, \chi; \epsilon)+ \epsilon\rho^{m\neq 0}_{rs}(a_r, a_s, \chi; \epsilon)\right), 
\end{align}
where all the $\sigma$ contain the pole structures while all the $\rho$ are regular for $\epsilon\to 0$ and vanish for $k^2\to 0$. For these reasons, we can evaluate all the $\rho$ at $\epsilon =0$. 
Specifically, we have:
\begin{subequations}
\begin{align}
    \sigma_{rr}(a_r,\epsilon)=& 4 \left[1- \frac{\Gamma(1+\epsilon)\Gamma(1-2\epsilon)}{\Gamma(1-\epsilon)}\left(\frac{1-a_r^2}{4}\right)^{-\epsilon}\right] \\
    \rho_{rr}(a_r;0)=& -\frac{1}{a_r}\Bigg[\left(a_r^2+4a_r-5\right)\ln \left(\frac{1-a_r^2}{4}\right)+ 2(5-a_r^2) \ln\left(\frac{a_r+1}{2}\right)\Bigg], 
\end{align}
\end{subequations}
\begin{subequations}
\begin{align}
    \sigma^{m=0}_{rs}(a_r,a_s,\chi;\epsilon)=&-2\left(\frac{1-a_r^2}{4}\right)^{-\epsilon}\frac{\Gamma(1-2\epsilon)\Gamma(1+\epsilon)}{\Gamma(1-\epsilon)}\left(1- 2\epsilon \frac{\beta_r}{\tanh\varphi}~_2F_1\left(1, \frac{1}{2}-\epsilon, \frac{3}{2},\beta_r^2\right)\right) \nonumber \\
    &+2 +(r\leftrightarrow s),\\
    \rho^{m=0}_{rs}(a_r,a_s,\chi;0)=& - \frac{2}{a_r} \ln\left(\frac{1+a_r}{2}\right)- \left(2-\frac{1}{a_r}\right)\ln\left(\frac{1-a_r^2}{4}\right)\nonumber  \\
    &-2\frac{1+ a_r a_s \cos\chi}{a_r^2+ a_r a_s \cos \chi}\xi_{rs}\tanh^{-1}(\xi_{rs})+ (r\leftrightarrow s), 
\end{align}
\end{subequations}
\begin{subequations}
\begin{align}
    \sigma^{m\neq0}_{rs}(a_r,a_s,\chi;\epsilon)=& - \left(\frac{1-a_r^2}{4}\right)^{-\epsilon} \frac{\Gamma(1-2\epsilon)\Gamma(1+\epsilon)}{\Gamma(1-\epsilon)}\left(1+2\epsilon \frac{\beta_r}{\tanh\varphi}~ _2F_1\left(1, \frac{1}{2}-\epsilon, \frac{3}{2},\beta_r^2\right)\right) \nonumber \\
    &+ \left(\frac{1-a_s^2}{4}\right)^{-\epsilon} \frac{\Gamma(1-2\epsilon)\Gamma(1+\epsilon)}{\Gamma(1-\epsilon)}\left(1-2\epsilon \frac{\beta_s}{\tanh\varphi}~ _2F_1\left(1, \frac{1}{2}-\epsilon,\frac{3}{2}, \beta_s^2\right)\right), \\
    \rho^{m\neq0}_{rs}(a_r,a_s,\chi;0)&= \frac{ a_s \cos \chi -5a_r}{a_r^2} \tanh^{-1}(a_r)+ 2D_{rs}\frac{\tanh^{-1}(\sqrt{\gamma_{rs}+ a_r^2(1-\gamma_{rs})})}{\sqrt{\gamma_{rs}+ a_r^2(1-\gamma_{rs})}} \nonumber \\
    &+2D_{sr}\frac{\tanh^{-1}(\sqrt{\gamma_{sr}+ a_s^2(1-\gamma_{sr})})}{\sqrt{\gamma_{sr}+ a_s^2(1-\gamma_{sr})}} - \ln\left(\frac{1-a^2_r}{1-a_s^2}\right) -\frac{a_s \cos \chi-a_r}{a_r},
\end{align}
\end{subequations}
where we have defined the following auxiliary parameters:
\begin{subequations}\label{eq: extra parameters}
\begin{align}
    \beta_r =& \frac{1}{\tanh\varphi}\left(1-\frac{1-a_r^2}{1-a_r a_s \cos\chi}\right), \quad&&\xi_{rs}= \frac{a_r^2+a_r a_s \cos\chi}{\sqrt{a^2_r+a_s^2+2a_r a_s \cos \chi-a_r^2a_s^2 \sin^2\chi}}, \\
    D_{rs}=& \frac{1+ a_r a_s \cos\chi}{a_r^2+ a_r a_s \cos \chi}\xi^2_{rs}, \quad &&\gamma_{rs}= \frac{-a_r^2a_s^2 \sin^2\chi}{a^2_r+a_s^2+2a_r a_s \cos \chi-a_r^2a_s^2 \sin^2\chi}.
\end{align}
\end{subequations}
We note that the $k^2\to 0$ limit is smooth for all the $\rho$ terms, whereas the $\sigma$ terms require a more careful treatment, as the limit must be taken at the level of distributions, in conjunction with the $\epsilon\to 0$ expansion of the factor $(k^2)^{-1-\epsilon}$.

\section{Details of the calculation:  integral over invariant mass}\label{app:calculations}
In this section we report the calculation of the double differential soft function defined in Eq.~\eqref{eq:massive_radiation_density_definition} for $\zeta=y$. As already stated, the following results are valid for any function of the ratio $k_+/k_-$. 
\subsection{Virtual correction}
The virtual correction to the double differential soft function is given by the integral of the one loop correction to the tree level eikonal current:
\begin{align}
\label{eq: massive virtual}
    &\frac{\de^2 \mathcal{S}_{\mathcal{Q},\text{V}}}{\de  \kt^2 \de y}= \frac{\mathcal{N}_\epsilon(\kt^2)}{2(2\pi)^{d-1}}\int^1_0 \frac{\de t}{t^2} \delta(1-t) |J_{\mathcal{Q}}(k(t,\kt,y))|_{\text{NLO}}^2= \left(\frac{\as(\mu^2)}{\pi}\right)^2 \frac{\cf 
    c(\epsilon)}{4} \left(\frac{\mu^2}{\kt^2}\right)^{\epsilon}\nonumber 
    \\
   &  \left(\frac{2p_1\cdot p_2}{(p_1\cdot p_k)(p_2\cdot p_k)}-\frac{m_1^2}{(p_1\cdot p_k)^2}-\frac{m_2^2}{(p_2\cdot p_k)^2}\right) \left[\ca\sum_{j=-2}^{0}r_j(\varphi,\alpha_{1k},\alpha_{2k})\epsilon^j
 \left(\frac{\mu^2}
 {\omega_k^2}\right)^\epsilon
 -\frac{\pi\beta_0}{\epsilon}\right]
\end{align}
where:
\begin{subequations} \label{eq:alpha-coeff}
    \begin{align}
    \alpha_{1k} =& \frac{m_1^2 (p_2\cdot p_k)}{(p_1\cdot p_2)(p_1\cdot p_k)}= \frac{\cosh\left(y+ \frac{\varphi}{2}\right)}{\cosh\left(y- \frac{\varphi}{2}\right)\cosh\varphi}, \\
\alpha_{2k} =& \frac{m_2^2 (p_1\cdot p_k)}{(p_1\cdot p_2)(p_2\cdot p_k)} = \frac{\cosh\left(y- \frac{\varphi}{2}\right)}{\cosh\left(y+ \frac{\varphi}{2}\right)\cosh\varphi}  \\
\omega^2_k =& \frac{2(p_1\cdot p_k)(p_2 \cdot p_k)}{(p_1\cdot p_2)} = \frac{\kt^2}{\cosh \varphi}\left(\cosh 2y+\cosh\varphi\right)
\end{align}
\end{subequations}
and the coefficients $r_{j}, j=-2,-1,0$ read:
\begin{subequations}
\begin{align}
    r_{-2}=& -\frac{1}{2},\\
    r_{-1}=& \frac{\varphi(1-\coth\varphi)+\ln\left(1+e^{-2\varphi}\right)}{2} \\
    r_{0}=& \frac{\zeta_2}{2}\left(\frac{19}{2}- 7\coth\varphi\right)- \frac{\varphi}{2}(1+\coth\varphi)\ln(1+e^{-2\varphi})- \frac{1}{4}\ln^2(1+ e^{-2\varphi}) \nonumber \\
    &+\frac{4\varphi\coth\varphi(-\alpha_{1k}+\alpha_{2k})\ln \frac{\alpha_{1k}}{\alpha_{2k}}+(\alpha_{1k}+\alpha_{2k})(4\varphi^2+ \ln^2 \frac{\alpha_{1k}}{\alpha_{2k}})}{16(\alpha_{1k}+\alpha_{2k}-2)}-\varphi\ln \left(\tanh\varphi\right)\coth\varphi \nonumber \\
    &+\frac{1}{2} \text{Li}_2\left(e^{-2\varphi}\right)\coth\varphi.
\end{align}
\end{subequations}
In the massless limit, i.e.\ $\varphi \to \infty$, the coefficient $r_{-1}$ vanishes, while $   \lim_{\varphi\to \infty} r_0= \frac{5}{4}\zeta_2$. Thus, the virtual corrections in the small mass limit read
\begin{align}
\label{eq: massive virtual small mass}
  \lim_{\varphi\to \infty}  \frac{\de^2 \mathcal{S}_{\mathcal{Q},\text{V}}}{\de  \kt^2 \de y}&=  \left(\frac{\as(\mu^2)}{\pi}\right)^2 \cf \ca
   \left(\frac{\mu^2}{\kt^2}\right)^{\epsilon} 
    \frac{2p_1\cdot p_2}{(p_1\cdot p_k)(p_2\cdot p_k)} \nonumber\\& \quad \quad \left[-\frac{1}{8 \epsilon^2} \left(\frac{\mu^2}{\kt^2} \right)^\epsilon-\frac{\pi \beta_0}{4 \ca \epsilon}+\frac{3}{8}\zeta_2+\ord{\epsilon}\right].
\end{align}
As discussed in Sec.~\ref{sec:massive_emitters}, the single massless limit is more delicate: we still have to take $\varphi\to \infty$ but keeping $y-\frac{\varphi}{2}$ fixed. This way, among the coefficients defined in Eq.~\eqref{eq:alpha-coeff}, $\alpha_{2k}$ vanishes while $\alpha_{1k}$ does not.

\subsection{Double-gluon emission}
In this section we report the calculation of the double differential soft function related to the double gluon emission. We express the result following the decomposition outlined in the previous section. Specifically we have to perform the calculation of the following integrals:
\begin{subequations}   \label{eq: final integrals}
    \begin{align}
    \frac{\de^2 \widetilde{\mathcal{S}}_{\mathcal{Q},rr}}{ \de  \kt^2 \de y} =&\left(\frac{\as(\mu^2)}{\pi}\right)^2 \left(\frac{\mu^2}{\kt^2}\right)^{2\epsilon}\frac{\ca \cf c(2\epsilon)}{16 \pi} \int^1_0 \frac{\de t}{t} \frac{m_r^2}{(p_r\cdot k)^2} \frac{t^\epsilon}{(1-t)^{1+\epsilon}}f_{rr}(a_r;\epsilon),    \label{eq: final integrals rr}  \\
    \frac{\de^2 \widetilde{\mathcal{S}}^{m=0}_{\mathcal{Q},rs}}{ \de  \kt^2 \de y} =&\left(\frac{\as(\mu^2)}{\pi}\right)^2 \left(\frac{\mu^2}{\kt^2}\right)^{2\epsilon}\frac{\ca \cf c(2\epsilon)}{16 \pi}\int^1_0 \frac{\de t}{t} \frac{p_r \cdot p_s}{(p_r \cdot k)(p_s \cdot k)}\frac{t^\epsilon}{(1-t)^{1+\epsilon}}f^{m=0}_{rs}(a_r, a_s,\chi; \epsilon),   \label{eq: final integrals rsm0}\\
    \frac{\de^2 \widetilde{\mathcal{S}}^{m\neq 0}_{\mathcal{Q},rs}}{ \de  \kt^2 \de y} =&\left(\frac{\as(\mu^2)}{\pi}\right)^2 \left(\frac{\mu^2}{\kt^2}\right)^{2\epsilon}\frac{\ca \cf c(2\epsilon)}{16 \pi}\int^1_0 \frac{\de t}{t} \frac{m_r^2}{(p_r \cdot k)^2}\frac{t^\epsilon}{(1-t)^{1+\epsilon}}f^{m\neq 0}_{rs}(a_r, a_s,\chi; \epsilon)   \label{eq: final integralsr rsm},
\end{align}
\end{subequations}
so that:
\begin{align}\label{eq:soft-function-S-tilde}
\frac{\de^2\widetilde{\mathcal{S}}_{\mathcal{Q}}}{\de \kt^2 \de y}= 2\frac{\de^2\widetilde{\mathcal{S}}^{m= 0}_{\mathcal{Q},12}}{\de \kt^2 \de y}+2\frac{\de^2 \widetilde{\mathcal{S}}^{m\neq 0}_{\mathcal{Q},12}}{\de \kt^2 \de y}+2\frac{\de^2 \widetilde{\mathcal{S}}^{m\neq 0}_{\mathcal{Q},21}}{\de \kt^2 \de y}-\sum_{r=1,2}\frac{\de^2\widetilde{\mathcal{S}}_{\mathcal{Q},rr}}{\de \kt^2 \de y}
\end{align}
We further observe that fixing the rapidity or any other function of the ratio $k_+/k_-$ of the off-shell emission make the integral over $t$ simpler, as in this case the parameter $\beta_r$ defined in Eq.~\eqref{eq: extra parameters} does not depend on $t$.

\subsubsection{The $\widetilde{\mathcal{S}}_{\mathcal{Q},rr}$ contribution}
Exploiting the considerations done in the previous paragraphs, it is easy to compute the integral in Eq.~\eqref{eq: final integrals rr}:
\begin{align}
\frac{\de^2 \widetilde{\mathcal{S}}_{\mathcal{Q},rr}}{\de \kt^2 \de y}=&\left(\frac{\as(\mu^2)}{\pi}\right)^2 \cf \ca \frac{m_r^2}{(p_r\cdot p_k)^2}\nonumber\left(\frac{\mu^2}{\kt^2}\right)^{2\epsilon} 
\Bigg[\frac{1}{4\epsilon^2}\left(1- \frac{1}{2}\left(\frac{\kt^2}{\omega^2_k}\right)^{-\epsilon}\right)\\
&+\frac{1}{8\epsilon}\left(\frac{\kt^2}{\omega^2_k}\right)^{-\epsilon} \ln \frac{\alpha_{rk}}{2}-\frac{1}{16}\ln^2\frac{\alpha_{rk}}{2}- \frac{\zeta_2}{2} 
   +\frac{1}{8}\Bigg(4\ln\left(\frac{1+u_r}{2}\right) \ln\left(\frac{1-u_r}{2}\right)\nonumber  \\
   &-(1-u_r)\ln\left(\frac{1-u_r}{2}\right)-(1+u_r)\ln\left(\frac{1+u_r}{2}\right)\Bigg)\Bigg],
\end{align}
where: 
\begin{align}
    u_1 = \sqrt{1-2\alpha_{1k} \frac{\kt^2}{\omega^2_k}}= \left|\tanh\left(y- \frac{\varphi}{2}\right)\right|, \quad u_2 = \sqrt{1-2\alpha_{2k} \frac{\kt^2}{\omega^2_k}}= \left|\tanh\left(y+ \frac{\varphi}{2}\right)\right|.
\end{align}

\subsubsection{The $\widetilde{\mathcal{S}}^{m=0}_{\mathcal{Q},rs}$ contribution}
The integration of $\widetilde{\mathcal{S}}^{m=0}_{rs}$ is more intricate due to the expression of $\rho^{m=0}_{rs}$, while the integration of $\sigma^{m=0}_{rs}$ is very simple. In order to expand the hypergeometric functions we use the \mathematica package HypExp \cite{Huber:2005yg},
\begin{align}
\label{eq: int S ij m=0}
\frac{\de^2 \widetilde{\mathcal{S}}^{m=0}_{\mathcal{Q},rs}}{\de  \kt^2 \de y}=&\left(\frac{\as(\mu^2)}{\pi}\right)^2\cf \ca \frac{p_r \cdot p_s}{(p_r\cdot p_k)(p_s\cdot p_k)} \left(\frac{\mu^2}{\kt^2}\right)^{2\epsilon} \nonumber \\
    \times&\Bigg[\frac{1}{4\epsilon^2}\left(1-\frac{1}{2}\left(\frac{\kt^2}{\omega^2_k}\right)^{-\epsilon}\right)+ \frac{1}{8\epsilon}\left(\frac{\kt^2}{\omega^2_k}\right)^{-\epsilon} \left(\varphi +\varphi \coth\varphi-\ln(1+ e^{2\varphi})\right)-\frac{\zeta_2}{2}\nonumber \\
    -& \frac{\coth\varphi}{8} \left(\ln \frac{\alpha_{rk}}{2} \tanh^{-1}(\beta_{rk})+\ln \frac{\alpha_{sk}}{2} \tanh^{-1}(\beta_{sk})+ \frac{G_{\mp}(\beta_{rk})}{2}+ \frac{G_{\mp}(\beta_{sk})}{2}\right)  \nonumber \\
    +&\frac{1}{32}\left(2\ln \alpha_{rk} \ln \alpha_{sk}- \ln^2 \left(2\cosh^2\varphi\right)-\ln^2 2\right)
    - \frac{1}{16}\int^1_0 \frac{\de t}{1-t} \rho^{m=0}_{rs}(a_r,a_s, \chi;0)\Bigg],
\end{align}
where we have defined:
\begin{align}
    \beta_{1k}= \frac{1-\alpha_{1k}}{\tanh\varphi}=- \tanh\left(y-\frac{\varphi}{2}\right), \quad   \beta_{2k}= \frac{1-\alpha_{2k}}{\tanh\varphi}=\tanh\left(y+\frac{\varphi}{2}\right)
\end{align}
The function $G_{\mp}(x)$ is a combination of
weight two generalised polylogarithms (GPLs)\cite{Goncharov:1998kja}.
Specifically, we use the standard convention:
\begin{align}
    G(a_1,\dots,a_n;z)= \int^z_0 \frac{\de t}{t-a_1}G(a_2,\dots,a_n;t), \quad G(;z)=1,
\end{align}
where as usual:
\begin{align}
    G(\underbrace{0,\dots,0}_{n};z)= \frac{1}{n!} \ln^n z; \quad G(\underbrace{0,\dots,0}_{n-1},1;z)= -\text{Li}_n(z), \quad G(\underbrace{a,\dots,a}_{n};z)= \frac{1}{n!}\ln^n(1-z/a).
\end{align}
The function $G_{\mp}(x)$ is given by:
\begin{align}
    G_{\mp}(x)= G(1,1;x)-G(1,-1;x)+G(-1,1;x)-G(-1,-1;x);
\end{align}
The left over integral reads:
\begin{align}
\label{eq: int rho}
-&\int^1_0 \frac{\de t}{1-t} \rho^{m=0}_{rs}(a_r,a_s,\chi;0)=2\ln\left(\frac{1+u_r}{2}\right) \ln\left(\frac{1-u_r}{2}\right)
+2\ln\left(\frac{1+u_s}{2}\right) \ln\left(\frac{1-u_s}{2}\right)
 \nonumber \\
&+4 \ln A \ln (1-A)+2(\coth\varphi-1)\left(G\left(\frac{1}{1-e^{-\varphi}},1,A\right)+G\left(-\frac{e^{-\varphi}}{1-e^{-\varphi}},0,A\right)\right) \nonumber  \\
&-2(1+\coth\varphi)\left(G\left(\frac{1}{1-e^{-\varphi}},0,A\right)+G\left(-\frac{e^{-\varphi}}{1-e^{-\varphi}},1,A\right)\right) \nonumber 
\\
&+ \left(G\left(\frac{1}{1-e^{-\varphi}},A\right)+G\left(-\frac{e^{-\varphi}}{1-e^{-\varphi}},A\right)\right)^2-2 G\left(1, \frac{1}{1-e^{-\varphi}}, A\right)\nonumber 
\\
&-2 G\left(0, \frac{1}{1-e^{-\varphi}}, A\right)-2 G\left(1, -\frac{e^{-\varphi}}{1-e^{-\varphi}}, A\right)-2 G\left(0, -\frac{e^{-\varphi}}{1-e^{-\varphi}}, A\right),\nonumber  \\
\end{align}
where:
\begin{align}
  A= \frac{1}{2}\left(1-\sqrt{1- \frac{2\kt^2}{\kt^2+(\omega_k^2-\kt^2)\cosh\varphi}}\right) = \frac{1}{2}(1-|\tanh y|).
\end{align}
It is interesting to study the small mass limit of Eq.~\eqref{eq: int S ij m=0}.
In this limit, we obtain:
\begin{align}
\label{eq: Sij m=0 masslessl limit}
   \lim_{\varphi\to\infty} \frac{\de^2 \widetilde{\mathcal{S}}^{m=0}_{\mathcal{Q},rs}}{\de  \kt^2 \de y}= \left(\frac{\as(\mu^2)}{\pi}\right)^2\cf \ca \frac{p_r \cdot p_s}{(p_r\cdot p_k)(p_s\cdot p_k)} \left(\frac{\mu^2}{\kt^2}\right)^{2\epsilon}\Bigg[\frac{1}{8\epsilon^2} - \frac{\zeta_2}{2}\Bigg].
\end{align}
If we perform the sum over the $(r,s)$ indices in Eq.~\eqref{eq: Sij m=0 masslessl limit} and then combine the result with the massless limit of the virtual corrections given in Eq.~\eqref{eq: massive virtual small mass}, we find the behaviour of the function $F$ quoted in Eq.~\eqref{eq:massive_FA_massless_limit}.

\subsubsection{The $\widetilde{\mathcal{S}}^{m\neq0}_{\mathcal{Q},rs}$ contribution}

Finally we examine the $m\neq 0$ contribution. We get:
\begin{align}
\frac{\de^2 \widetilde{\mathcal{S}}^{m\neq0}_{\mathcal{Q},rs}}{\de  \kt^2 \de y}=&  \left(\frac{\as(\mu^2)}{\pi}\right)^2\frac{ m_r^2}{(p_r \cdot p_k)^2} \cf \ca \left(\frac{\mu^2}{\kt^2}\right)^{2\epsilon}\nonumber \\
&\Bigg[\left(\frac{\kt^2}{\omega_k^2}\right)^{-\epsilon}\frac{1}{16\epsilon}\left(-\varphi \coth\varphi+ \frac{1}{2}\ln \frac{\alpha_{rk}}{\alpha_{sk}}\right)- \frac{\ln \frac{\alpha_{rk}}{4}\ln \alpha_{rk}-\ln \frac{\alpha_{sk}}{4}\ln \alpha_{sk}}{64} \nonumber \\
&  + \frac{\coth\varphi}{32} \left(G_{\mp}(\beta_{rk})+G_{\mp}(\beta_{sk})+ 2\tanh^{-1}\beta_{rk}\ln \frac{\alpha_{rk}}{2}+2\tanh^{-1}\beta_{sk}\ln \frac{\alpha_{sk}}{2}\right)\nonumber \\
&-\frac{1}{16}\int^1_0 \frac{\de t}{1-t} \rho_{rs}^{m\neq 0}(a_r, a_s, \chi; 0)\Bigg].
\end{align}
The integral of $\rho^{m\neq 0}$ is given by: 
\begin{align}
    -\int^1_0 \frac{\de t}{1-t} \rho^{m\neq 0}_{rs}&= \frac{2}{\tanh\varphi}\bigg[
  G\Big(-1-\tanh\varphi,0,\frac{1-B}{B}\Big)
- G\Big(-1+\tanh\varphi,0,\frac{1-B}{B}\Big)\nonumber 
\\
&+ \ln(\cosh\varphi)\bigg(\ln\Big(\frac{1+B\tanh\varphi}{1-B\tanh\varphi}\Big)-2\varphi\bigg)\bigg]
+ \ln^2\!\Big(\frac{(1-B)\cosh\varphi}{B}\Big)\nonumber\\
&
+ \frac{1}{4}\ln^2\!\Big(\frac{\alpha_{rk}}{\alpha_{sk}}\Big)
- \frac{1}{2}\ln^2\!\Big(\frac{1-u_{r}^2}{4}\Big)
- \frac{1}{2}\ln^2\!\Big(\frac{1-u_{s}^2}{4}\Big)
+ 4\ln\Big(\frac{1+u_{r}}{2}\Big)\ln\Big(\frac{1-u_{r}}{2}\Big)
\nonumber\\
&+ \frac{1-\alpha_{rk}}{\alpha_{rk}}\bigg(\ln\Big(\frac{1-u_{r}^2}{4}\Big)+\frac{2\tanh^{-1}u_{r}}{u_{r}}\bigg)
\end{align}
where we have defined:
\begin{align}
    B= \frac{\kt^2}{\omega_k^2-\sqrt{\omega_k^4-2\kt^2 \omega_k^2+ \kt^4 \tanh^2\varphi}}= \frac{\cosh\varphi}{\cosh2y+\cosh\varphi- |\sinh2y|}
\end{align}

\subsection{Final result for the soft function}

\begin{figure}
    \centering
  \includegraphics[width=1.0\linewidth]{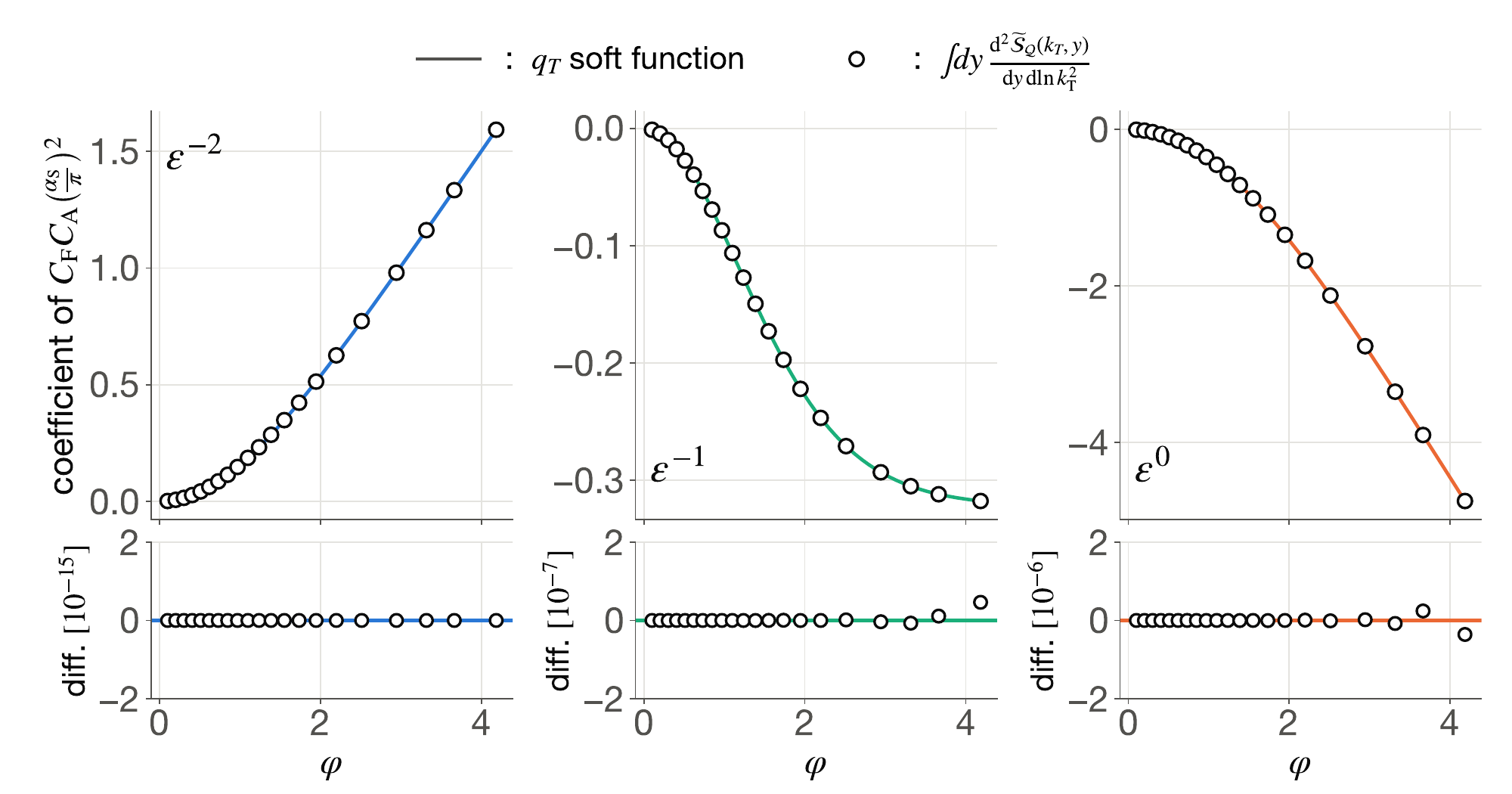}
    \caption{Check of the integrated double pole, single pole, and finite part of the effective coupling against the results of \cite{Catani:2023tby,Devoto:2025eyc}.}
    \label{fig:shark-checks}
\end{figure}

The soft function presented in~\cite{Catani:2023tby, Devoto:2025eyc} can be expressed in terms of the same integrals we presented in this paper for the calculation of the effective coupling, after a further integration over the transverse momentum $\kt$.
We have thus performed a further cross check of our expressions by integrating numerically.
Since the additional integration leads to a further pole in $\epsilon$, the one-to-one correspondence of the integrals gets shifted by one order, and our finite remainder has to be compared against the single pole of~\cite{Catani:2023tby, Devoto:2025eyc}.
We find excellent agreement between the two independent computations, as demonstrated in Fig.~\ref{fig:shark-checks}.

Finally,
the full result for the soft function at $\ord{\as^2}$ is obtained by adding to the eikonal result Eq.~\eqref{eq:massive_radiation_density_LO} 
the $\widehat{S}$ and $n_f$ contributions Eq.~\eqref{eq:soft-function-S-hat} and~\eqref{eq:massive_nf_fixed_y}, together with the correlated real-emission contributions of Eq.~\eqref{eq:soft-function-S-tilde}, and the remaining parts of the virtual corrections, namely the non-$\beta_0$ terms in Eq.~\eqref{eq: massive virtual}:
\begin{equation}
   \frac{\de^2 \mathcal{S}_{\mathcal{Q}}}{\de  \kt^2 \de y}=   
\frac{\de^2{\mathcal{S}}_{\mathcal{Q},\text{LO}}}{\de \kt^2 \de y}
+
\frac{\de^2\widehat{\mathcal{S}}_{\mathcal{Q}}}{\de \kt^2 \de y}+
\frac{\de^2{\mathcal{S}}_{\mathcal{Q},n_f}}{\de \kt^2 \de y}
+\frac{\de^2\widetilde{\mathcal{S}}_{\mathcal{Q}}}{\de \kt^2 \de y}+
\frac{\de^2{\mathcal{S}}_{\mathcal{Q},\text{V}\neq \beta_0}}{\de \kt^2 \de y}.
\end{equation}
In the above sum, all $1/\epsilon$ poles cancel and the combinations of GPLs that appear in the 
finite remainder can be substantially simplified.
The resulting expression can be cast in the form presented in Eq.~\eqref{eq:massive_fixed_y_result}.

\bibliography{massCMW}
\end{document}